\documentclass[12pt]{article}

\usepackage[dvips]{graphics}
\usepackage{rotating}
\usepackage{epsfig}
\def\lesssim{\mathrel{\hbox{\rlap{\hbox{\lower5pt\hbox{$\sim$}}}\hbox{$<$}}}}
\def\gtrsim{\mathrel{\hbox{\rlap{\hbox{\lower5pt\hbox{$\sim$}}}\hbox{$>$}}}}

\def\ntrli{\chi^0_i}

\def\chipm{\chi_i^\pm}
\def\chjpm{\chi_j^\pm}

\def\chip{\chi_i^+}

\def\chim{\chi_i^-}

\newcommand{\ntrl}[1]{\chi^0_#1}
\newcommand{\chpm}[1]{\chi^\pm_#1}

\def\sleppm{\tilde{\ell^\pm}}   %

\def\snu{\tilde{\nu}}

\def\squark{\tilde{q}}
\def\squarkl{\tilde{q}_L}

\newcommand{\sstop}[1]{\tilde{t}_#1}
\def\gluino{\tilde{g}}

\def\ellpm{\ell^\pm}       %
\def\ellbar{\bar{\ell}}       %
\def\ellbarprime{\bar{\ell}^\prime}       %
\def\qprime{q^\prime}       %
\def\qbar{\bar{q}}       %
\def\qbarprime{\bar{q}^\prime}       %
\def\mone{M_1}
\def\mtwo{M_2}
\def\mthree{M_3}
\newcommand{\mntrl}[1]{m_{\chi^0_#1}}      %
\newcommand{\mchpm}[1]{m_{\chi^\pm_#1}}

\def\mslep{m_{\tilde{\ell}}}
\def\mslepl{m_{\tilde{\ell}_L}}
\def\mslepr{m_{\tilde{\ell}_R}}

\def\msnu{m_{\tilde{\nu}}}

\def\msquark{m_{\tilde{q}}}
\def\msquarkl{m_{\tilde{q}_L}}
\def\msquarkr{m_{\tilde{q}_R}}

\newcommand{\mstop}[1]{m_{\tilde{t}_#1}}
\def\mgluino{m_{\tilde{g}}}

\def\tanbeta{\tan\beta}

\def\ptjet{p_T^{jet}}

\def\ptlepton{p_T^{lepton}}

\def\etalepton{\eta_{lepton}}

\def\etmiss{\not\!\!{E_T}}

\def\fbinverse{{\mathrm{fb}}^{-1}}

\def\beq{\begin{equation}}   %
\def\eeq{\end{equation}}   %
\def\bea{\begin{eqnarray}}   %
\def\eea{\end{eqnarray}}   %
\def\non{\nonumber}

\begin{document}


\begin{flushright}
MCTP-05-95\\
October 2005
\end{flushright}
\vspace{2cm}

\begin{center}
{\Large\bf 

Is it SUSY?} \\
\vskip 15pt
{\large AseshKrishna Datta\footnote{asesh@umich.edu},  
Gordon L. Kane\footnote{gkane@umich.edu} and
Manuel Toharia\footnote{mtoharia@umich.edu}}  \\
\vskip 10pt
{\large Michigan Center for Theoretical Physics (MCTP) \\
University of Michigan, Ann Arbor, MI 48109, USA}
\end{center}

\vskip 65pt
\centerline{\large\bf Abstract}
\vskip 15pt
\noindent

If a signal for physics beyond the Standard Model is observed at the
Tevatron collider or LHC, we will be eager to interpret it. Because
only certain observables can be studied at a hadron collider, it will
be difficult or impossible to measure masses and spins that could
easily establish what physics was being seen. Nevertheless, different
underlying physics implies different signatures. We examine
the main signatures for supersymmetry, with some emphasis on
recognizing supersymmetry in parts of parameter space where generic
signatures are reduced or absent. We also consider how to distinguish
supersymmetry from alternatives that most closely resemble it, such as
Universal Extra Dimensions (UED). Using the robust connection between spins
and production cross section, we think it will not be difficult to
distinguish UED from supersymmetry. We expect that by considering
patterns of signatures it is very likely that it will not be difficult
to find a compelling interpretation of any signal of new physics.

\vspace{2cm}

\newpage
\section{Introduction}
An exciting era in particle physics will begin
with the commencement of the Large Hadron Collider (LHC) at CERN 
in about 2007. 
While we do not know for sure what is
waiting for us there, we do have a concrete wish list and
rather solid directions to follow, which evolved
from hints based on previous phenomena.
Understanding Electroweak
Symmetry Breaking (EWSB), especially in terms of the 
Higgs mechanism, is still the main priority.

A well known issue intimately related to the presence of a 
fundamental scalar in the Standard Model (SM) is the
hierarchy problem. New physics such as supersymmetry (SUSY) at the weak scale
has long been known to cure this problem at a
fundamental level. There is ample reason to expect a 
deeper connection between a weakly coupled Higgs-induced EWSB 
phenomenon and weak-scale SUSY. A large part of the
LHC program is based on this anticipation.
As a bonus, on the cosmological front, the lightest of the 
supersymmetric particles (the LSP) in a model with a conserved 
discrete symmetry ($R$-parity) provides us with a viable
dark matter candidate which is lacking in the SM. 

Efforts have been underway for 15-20 years
to study the collider signatures of different SUSY models at a 
theoretical level. Also, recent and ongoing experiments at CERN-LEP 
and the Fermilab Tevatron had/have dedicated programs to directly find
the SUSY partners of Standard Model particles.
Imprints of new kind of physics were/are also expected in numerous 
other experiments involving proton decay, electric and magnetic dipole moments,
rare decays, heavy-quarks (the $B$-factories) and dark-matter 
experiments. So far there is no unambiguous positive signal for SUSY 
from these experiments. For direct observations this is not
discouraging, since LEP and the Tevatron
with their low available center of mass energies and luminosities
could only probe the lower end of the SUSY spectrum. 
They already
did a good job of ruling out certain regions of the SUSY 
parameter space in direct searches 

Here, we revisit approaches to recognizing SUSY
in future data. There should be two intimately related but
potentially distinct programs: first,
to be confident that a signal is SUSY and nothing else and, second, to learn as
much as possible about the effective lagrangian, which should provide
hints about how supersymmetry is broken, and about the underlying theory.
Of course, there are a number of signals that would be inconsistent
with an R-parity conserving SUSY interpretation, such as wide WW or
jet-jet resonances, or no missing transverse energy.

The most characteristic feature of the SUSY partners 
for SM particles is their spin:
\emph{i.e.}, spin `0' (scalar) partners for every known 
(spin 1/2) SM fermions 
and partner fermions for spin `1' gauge bosons of the SM. 
However measuring spins of newly observed particles 
at the LHC has not been studied until very recently 
\cite{Barr:2004ze} (see also \cite{Smillie:2005ar} and \cite{Datta:2005zs}).

Indeed, except in very special cases, it will not be possible to
isolate individual candidate superpartners at all, and measure their masses
or spins, because of the nature of hadron collider data. It will also
be very hard to test whether electroweak superpartners have correct
gauge quantum numbers and couplings. 


In this paper we will study both typical and less typical observables
that are expected if the new physics is supersymmetry. We examine in
some detail what happens when some common leptonic channels are
suppressed, while other signatures remain robust. Such a situation is not commonly 
encountered in minimal Supergravity (mSUGRA) based `benchmark' 
scenarios which have been overwhelmingly used in mapping out the search 
strategies. It may be important to be aware of such possibilities 
beforehand. 
On the other hand, we point out that such low event rates 
may help decipher the underlying spectrum.

Then we examine what a signal could be if it is not SUSY. 
We will discuss the example most often proposed as difficult to
distinguish from SUSY, universal extra dimensions (UED), and study a little
how to distinguish it with LHC observables.
We propose one method, based on the robust relation between spins and
production cross sections, of distinguishing SUSY and UED (and other
models). Essentially, once there is data to input to an analysis,
cross sections determine spins even at a hadron collider. We also note
that patterns of signatures for SUSY and for UED are rather distinguishable.

Throughout this work special reference is made to events with 
Same-Sign (known also as Like-Sign) Dilepton (SSD). These events
have been long known 
[1-10]
to be very clean and 
robust signatures at the hadron colliders for physics Beyond the
Standard Model (BSM), because the Standard Model (SM) background for
the SSD events is known to be very small and can be effectively eliminated.
In the context
of supersymmetry, the presence of a heavy majorana fermion like the 
gluino (the superpartner of the SM gluon) is a natural 
source of SSD.
In fact, gluinos give not only SSD's but also predict their rate to be 
equal to that for the Opposite-Sign Dilepton (OSD) pair [2-5].
This makes the SSD very characteristic of physics beyond the SM, 
particularly SUSY. As is expected,
extensive studies, some involving detailed simulation of this final state, 
are already present in the literature [4-10].
However, most of them addressed the SSD events along with 
other conventional leptonic and/or jets final states that arise in 
a specific SUSY scenario (like the minimal Supergravity (mSUGRA)) or in 
some limited corner of an otherwise more general
framework, like the Minimal Supersymmetric Standard Model (MSSM). 
Instead, in this work, our focus is primarily 
on the SSD events in the full MSSM.

While the SSD events potentially constitute ``smoking-gun'' signals for some non-standard 
physics, two pertinent issues need to be addressed carefully:
\begin{enumerate}
\item What could be the implications of observing non-SM signals but
few or no SSD's?
\item How far can they go in singling out a particular BSM scenario like SUSY; 
i.e. could there be more than one candidate scenario to explain the SSD events?
\end{enumerate}
In short, though wonderfully characteristic, how robust and useful 
are the SSD events for actual purposes?

In this paper we want to be as phenomenological as possible. The low
scale world could be supersymmetric but not described by the minimal
case, even the full MSSM. {\it All} usual constraints, such as
electroweak symmetry breaking, dark matter, $b\rightarrow s\gamma$,
etc., are significantly model dependent, so we should not and do not
impose any of them.

The paper is organized as follows. 
Our discussion will be in the context of a $pp$ collider 
like the upcoming Large Hadron Collider (LHC). Much of the discussion
is also relevant for the
Tevatron, but for simplicity we focus on the LHC. 
%
In Section 2 we discuss how the SSD could be produced and point out 
some associated features.  Basic requirements for having the SSD are sorted out
and we discuss to what extent these are unique and/or robust for SUSY scenarios. 
This leads to the question whether SUSY can thrive without the SSD.
Section 3 introduces the set-up for our SUSY analysis, the general assumptions
and some tools.  
Event rates for the SSD along with other possible final states for 
representative MSSM spectra are studied. 
In Section 4 we critically review situations where SSD events are absent.
We discuss how such null-observations, when combined with a careful study of 
the other channels (inclusive studies), could provide  a wealth of informations 
on the underlying framework. 
In section 5 we clarify whether a different theoretical framework could possibly
lead to very similar observations and thus could fake SUSY. 
We then conclude.
%
\section{The Same Sign Dileptons: Origin, Anatomy and the Caveats}
\subsection{The Origin}
At hadron colliders, the source of SSD originally discussed [2-5] was 
the pair-production of gluinos followed by their 
leptonic decays (assuming $\msquark > \mgluino$), 
\bea
\mbox{\large $ \mathop{}_{\rm\textstyle pp}^{\rm\textstyle p\bar{p}} \longrightarrow\tilde{g}$}
\!\left(
\stackrel{\tilde{q}^*}{\longrightarrow} q \bar{q}\  \tilde{\chi}^{\pm}_i
\!({\scriptstyle \stackrel{\tilde{\ell}^*}{\longrightarrow}\ \ell^{{}^\pm} \nu \chi^0_i})\
\right)\ + \ \
\mbox{\large $\tilde{g}$}
\!\left(
\stackrel{\tilde{q}^*}{\longrightarrow} q\bar{q}\ 
\tilde{\chi}^{\pm}_i
\!({\scriptstyle \stackrel{\tilde{\ell}^*}{\longrightarrow}\ \ell^{{}^\pm} \nu \chi^0_i})\
\right)\non.
\eea
Because of the majorana nature of the gluino\footnote{The authors of
\cite{Barnett:1988mx} used a broader definition of `Majorana' to include
neutral particles which transform under real representations of the
underlying SM gauge group. In Section 5 we discuss how this generalization 
works for an alternative scenario.}, it has an equal probability to decay 
into $\chip$ and $\chim$ \cite{Barnett:1988mx, barnett2}. This means
an equal rate for decay through an off-shell
$\squark$ or via its charge conjugate state. 
The sign of charge on
the leptons follows that of the chargino which, in turn, follows
the off-shell squark. Thus in half of the cases gluinos decay 
into same-sign charginos leading to SSD. 

The associated production of a squark and a gluino can lead to an SSD
event where the matching-charge lepton comes from the gluino decay, or
SSD events could even originate from less likely cascades of same
sign squarks, \emph{viz.}, 
\[ \squark \, \to \, \qprime \, \chipm \to \, \qprime \, \ellpm \, \nu \, \ntrli \] 
Also, the associated production of a squark and a gluino can lead to an SSD
event where the matching-charge lepton comes from the gluino decay; if
squarks are not heavy this is an important channel. 

Note that if the gluino or the squark is massive enough to 
decay into the charginos, these decays dominate over its decays into the 
neutralinos, \emph{i.e.}, $\gluino \to q \qbar \ntrli$ 
\cite{Baer:1986au, Barnett:1987kn, Bartl:1990ay}.

Whatever the relative masses of the squark and the gluino, effectively only
the left-handed
squarks take part in the EW cascade leading to the chargino when the latter is
gaugino-dominated ($\mu >> \mtwo$). When the chargino is higgsino-dominated
($\mu << \mtwo$) both do take part but the yukawa-like couplings are only 
proportional to the corresponding quark-masses.

Decays of heavier neutralinos (which are also majorana fermions) to charginos
and $W^\pm$ bosons followed by the leptonic decay of either of the latter
could also contribute to SSD events:
\[ \ntrli \to \chjpm W^\mp \]
These neutralinos could be directly produced in the hard scattering via
weak-interactions (in which case they are suppressed at the LHC)
or be present in the cascade of a squark or a gluino.

\subsection{The Anatomy}

Interestingly, cascades of pair-produced gluinos 
would lead not only to an equal number of SSD and OSD events but also to
the same number of $(++)$ and $(--)$ lepton-pairs. 
The majorana nature of the gluino
guarantees both. Thus, 
any significant charge-asymmetry in SSD events 
would mean they are not solely originating in gluino-cascades 
\cite{Baer:1991xs, Baer:1995va}. 

On the other hand, usually\footnote{SUSY cascades involving multiple top and
bottom quarks can do the job as well when one of the leptons come from a
leptonic $b$ quark decay. This is the same mechanism, albeit under a SUSY
cascade, that accounts for a reducible SM background to SSD events
\cite{Baer:1995va}.}
it takes a squark pair of the same sign when their cascades (involving 
charginos) lead to SSD (as is the case when $\mgluino >  \msquark$). 
When these squarks come from the strong decay of the heavier gluino
once again one expects an equality of of $(++)$ and $(--)$ pairs.
However, squarks produced directly would more likely be positively charged, 
the LHC being a $p\,p$ machine. 
Hence the LHC would see a charge-asymmetry in having 
more of $(++)$-SSD pairs compared to $(--)$ ones 
\cite{Baer:1991xs, Baer:1995va}. 
This would reflect on squark-gluino mass hierarchy 
with squarks being lighter than the gluino.

Also, the cascade-patterns discussed in Section 2.1 indicate that
the jet-multiplicity
is expected to be larger for gluino-cascades \cite{Baer:1995va} compared to 
squark-cascades. This could be another useful observable.

Hence, the typical source of an SSD event
is the availability of
two same-sign charginos at any stage of the decay chains (see footnote).
Note that parton-level hard scattering 
never leads to such a doubly-charged configuration 
in a two-body final state.

Association of the SSD events with majorana fermions,  
though very characteristic, may not be unique.
The necessary and an \emph{a priori} sufficient requirement 
is the presence of a ``chargino-like'' massive state in the spectrum 
with an electroweak 
quantum number 
(thus coupling to SM fermions and gauge bosons, see footnote 4 for a rather 
futuristic generalization in the original paper on this subject) 
which is
kinematically accessible for a cascade.  In section 5 we provide a 
scenario which gives SSD by exploiting this feature and does not contain any 
`majorana fermion'.
Until then we restrict ourselves to SUSY scenarios,
and the MSSM in particular.

\subsection{The Caveats}
Over the last two decades numerous studies have 
confirmed the connection of SUSY and SSD both theoretically and at the level of
simulations. We do not
know of any work where the contrary has been systematically addressed
before the present paper. 

As mentioned in the Introduction, one of the main purposes of the present 
work is to understand how necessary SSD's are if SUSY exists. With
an understanding of the origin of the SSD as discussed in 
last two sections we now describe situations where SUSY might not 
lead to SSD. A signal of new physics combined with the absence of SSD
in the initial LHC data from  
a low luminosity run could immediately turn attention
to some definite regions of the SUSY parameter space. In practice, 
indications for SUSY might start coming simultaneously from several
channels. 
However, given that the SSD events are exceptionally clean,
analyzing and interpreting them is likely to be
less complicated and hence, fast. 

We now briefly discuss some situations within the MSSM where the SSD signal 
would turn out to be rather feeble: 
\begin{itemize}
\item Few SSD events would result 
if the charginos are not kinematically accessible down the cascades.
This happens when the squarks or the gluino, whichever is lighter (and 
initiates the electroweak cascade), is almost degenerate with or lighter 
than the charginos.
\item When the gluino mass lies between the masses of different
  squarks, it is possible that the supersymmetric decays show a
  preference to non-leptonic channels. An example is when left handed
  squarks are heavier than the gluino, which in turn is heavier than
  right handed squarks. In this case left handed squarks, if directly
  produced, will preferably decay into gluinos. The main decay channel
  of these will be into quark and right handed squark, and then
  depending on the EW gaugino spectrum right handed squarks may decay
  mainly into quarks and LSP, thus eliminating  leptons from the SUSY cascade.
\item In the split SUSY scenarios 
\cite{Arkani-Hamed:2004fb, Giudice:2004tc, Arkani-Hamed:2004yi}
with very heavy sfermions, it was recently pointed out
\cite{Toharia:2005gm} that the 3-body decays of the gluino may become
marginal due to the presence of strong 2-body radiative decays
($\gluino \, \to \, g \ntrli$)
enhanced by large logs coming from the
higgsino content in the neutralino\footnote{A subsequent work \cite{Gambino:2005eh}
with improved treatments (resummed logs) confirms the trend (gluino 2-body
decays can overcome easily 3-body decays), although
the effect is smoothed out when the SUSY breaking scale becomes very
large in which case the log resummation becomes a necessary
computation.}.
Having a new decay channel for the gluino opens the possibility of
depleting leptons in new ways.
\end{itemize}
 
Then there are situations where SSD's are depleted as a result of
an interplay of the SUSY spectrum on the theoretical side
and the detector and the simulation
criteria employed on the experimental side. For example, 
\begin{itemize}
\item A wino-like lighter chargino and the LSP neutralino can be very 
degenerate in the Anomaly-Mediated SUSY Breaking (AMSB) scenarios
\cite{Randall:1998uk, Giudice:1998xp}. This renders 
the decay-lepton in  
$\chipm \, \to \, \ellpm \nu \ntrl1$
too soft to pass the usual lepton-trigger at the LHC.

\item In the ``effective'' (or ``virtual'') Lightest SUSY Particle 
  scenario (VLSP) \cite{Datta:1994un, Datta:1994ac, Datta:1996ur} 
the sneutrino mass lies between the mass of the two lightest
  neutralinos. All decays of the second lightest neutralino $\ntrl2$
  go into invisible neutrino-sneutrino pairs, thus eliminating a 
  possible source of leptons. When the mass difference between the
  sneutrinos and the lightest chargino $\chpm1$ is small enough, leptons
  produced in $\chpm1$ decays might be too soft to trigger on.  

\end{itemize}

\noindent
In the next two sections we take up all these issues systematically.

\section{Event Rates: the Usual paradigm}

Following the existing studies, the channels that we discuss in addition to 
SSD's are
(i) OSDs with two or more jets, (ii) 1-lepton + two or more jets, 
(iii) trilepton+jets and (iv) inclusive multi-jets ($n_{jets} \geq 3$).

We do not discuss various backgrounds as they are quite
generic, well-known and much studied. 
We always use the ATLAS \cite{atlas}
criterion for the definition of reach for the 
masses of the SUSY particles in different channels which looks for at least
10 events with $N_{signal}/\sqrt{N_{background}} > 5$ for an integrated 
luminosity of $10 \;\fbinverse$. In practice, we would assume 
the SSD events to be free from SM backgrounds thus looking for only 10 events
for it at 10 $\fbinverse$. Our main results are not sensitive to these criteria.

In the following subsections we outline the broad theoretical framework, 
the scope of the study and the set-up for the analysis followed by a selection 
of representative sets of MSSM parameters for demonstration.


\subsection{The Framework and the Set-up}
The popular phenomenological framework we adopt for this section is
broadly characterized by having 
the $U(1)$ and the $SU(2)$ gaugino masses lighter than the $SU(3)$ gaugino
mass at the weak scale, \emph{i.e.}, $\mone, \mtwo < \mthree$. 

We divide our study broadly into two cases: (1) $\msquark > \mgluino$ and 
(2) $\mgluino > \msquark$; which provide two functionally distinct sources
for SSD's-- the gluino in the first case and the squark in the latter, as
already discussed in section 2.1. 


We work at the parton-level, and do not consider
tau or bottom decays, or initial or final state radiation (unless
explicitly stated).
Neither do we consider beam-induced effects like beam remnants, minimum-bias
or pile-ups, and jets are pure partonic 
in nature, not treated with any jet-algorithm (jet-merger was not
done). 
We do not include any efficiency factor 
other than what
is already implied by the choice of triggers and the nominal kinematic 
cuts (used
by the LHC collaborations).
However, we make general comments on expectations under 
a full simulation. None of these effects would make qualitative
changes in our conclusions.
We generated events with {\tt Pythia v6.316} 
\cite{Sjostrand:2003wg} which uses the SUSY Les Houches Accord (sLHA) interface 
\cite{Skands:2003cj}
to a SUSY spectrum generator like {\tt SuSpect v2.33} \cite{Djouadi:2002ze}.
The parton distribution we use is CTEQ5L \cite{Lai:1999wy}
which is the default to 
{\tt Pythia v6.316}. We stick to this set-up throughout this work.

On the experimental side, we consider LHC-data 
expected from the first 
year of a low-luminosity run 
($\sim 10 \; {\mathrm {fb}}^{-1}$ over a year),
where a detailed study of the SSD's could prove specially informative. 
Nominal triggers and kinematic cuts used in the present analysis are motivated
by the ones used in the inclusive SUGRA studies  by the ATLAS Collaborations
\cite{atlas}. To be specific, these are: 

\[ \ptjet > 100 \; {\mathrm {GeV}}, \;\; \etmiss > 100 \; {\mathrm {GeV}}, \;\;
\ptlepton > 20 \; {\mathrm {GeV \; and}} \; \; \etalepton < 2.5\ . \]

\noindent
While we move on to slightly different theoretical framework(s) in the 
subsequent section(s), the set-up described above remains the same
throughout this work unless otherwise indicated.

\subsection{Choice of MSSM parameters and the Event Rates}
In this subsection we discuss the event rates for the relative hierarchies
between squark and the gluino masses as described above.
We set soft squark masses at the weak scale to be
degenerate, and the 
trilinear $A$ parameters are also taken to be degenerate and kept small 
so that the mixings in the sfermion sector are in turn small and no specific
phenomenology emerges dominantly from that sector. The three 
soft gaugino masses are defined at the weak scale with arbitrary
mutual ratios, as opposed to scenarios which presume
a gauge coupling unification at a high scale. We initially take $\mone <
\mtwo < \mthree$, and discuss later how things
change with different hierarchies. The higgsino
mass parameter $\mu$ and $\tanbeta$ are also input. As mentioned
above, we do not impose EWSB conditions since they do not hold in
extended MSSMs, and we want to study the parameter space more generally. 

The upper panel of Fig. 1 illustrates the cases for $\msquark > \mgluino$.
Br[$\squark \to q \gluino$] is 
almost 100\% and only electroweak cascades of gluinos matter. 
Situations with
(i) varying soft squark masses with a fixed gluino mass and
(ii) varying gluino masses with a fixed soft squark masses are presented
which reflect the trends over the squark-gluino mass plane
subject to their specific hierarchies. The other relevant
SUSY parameters are as indicated in the figure caption. 
The lower panel with a similar set of figures illustrates the reverse case, 
\emph{i.e.}, $\mgluino > \msquark$.

If squarks (gluino) are not too heavy, the production rates for
squarks (gluinos) contribute significantly 
to the cascades via the gluino (squark) decay to SSD events, as does the associated
production of gluinos and squarks. For a given $\mgluino(\msquark)$, 
the squark (gluino)
contribution to SSD events decreases gradually with increasing 
$\msquark (\mgluino)$.
As expected, with increasing mass, the
residual events 
in the inclusive signal come mainly from the production of lighter 
gauginos. Observation of such a kind of 
event-pattern 
would immediately indicate a decoupled spectrum for the squarks (sfermions in 
general) and the gluino while suggesting lighter gauginos.
It could also be suggestive of 
some kind of non-universality of gaugino masses at some high scale.

\begin{figure}[t]
\begin{center}
\centerline{
\hspace{-2.7cm}\epsfig{file=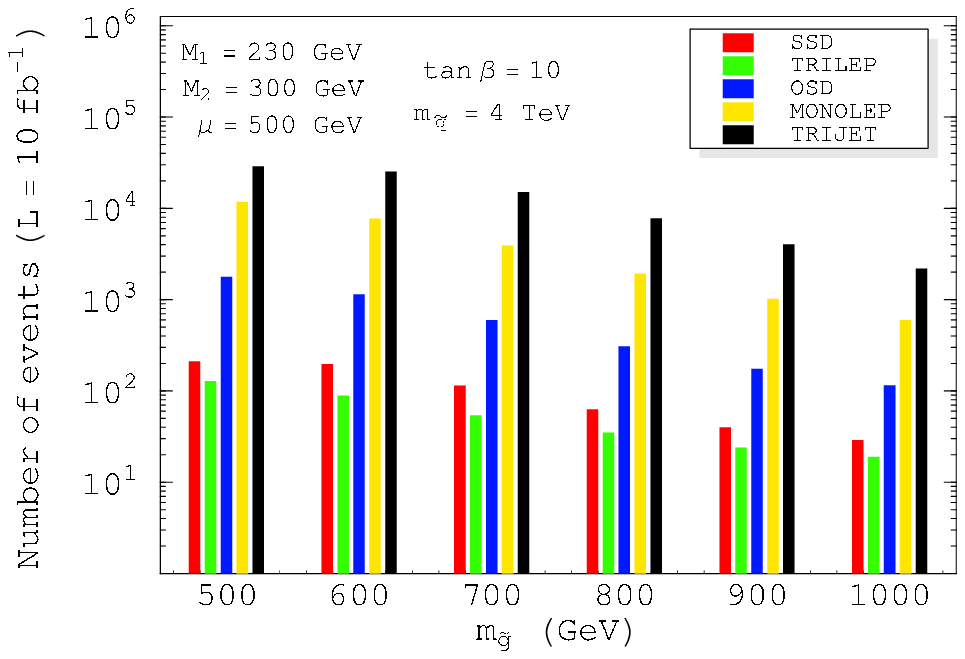,width=9.7cm}
\hspace{-1.7cm}
\epsfig{file=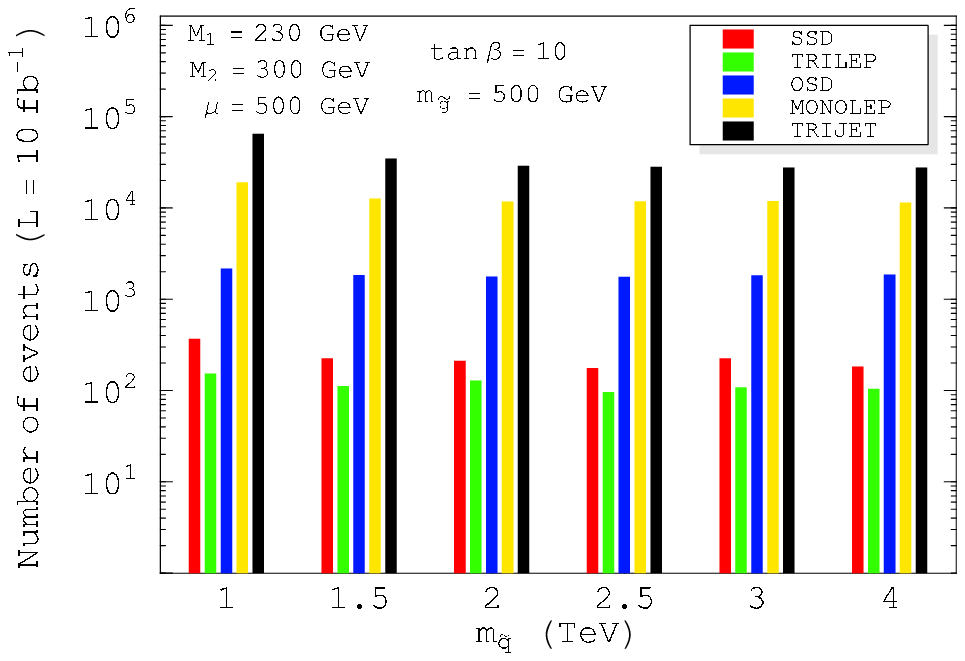,width=9.5cm}}
\hspace{-2.7cm}\centerline{\epsfig{file=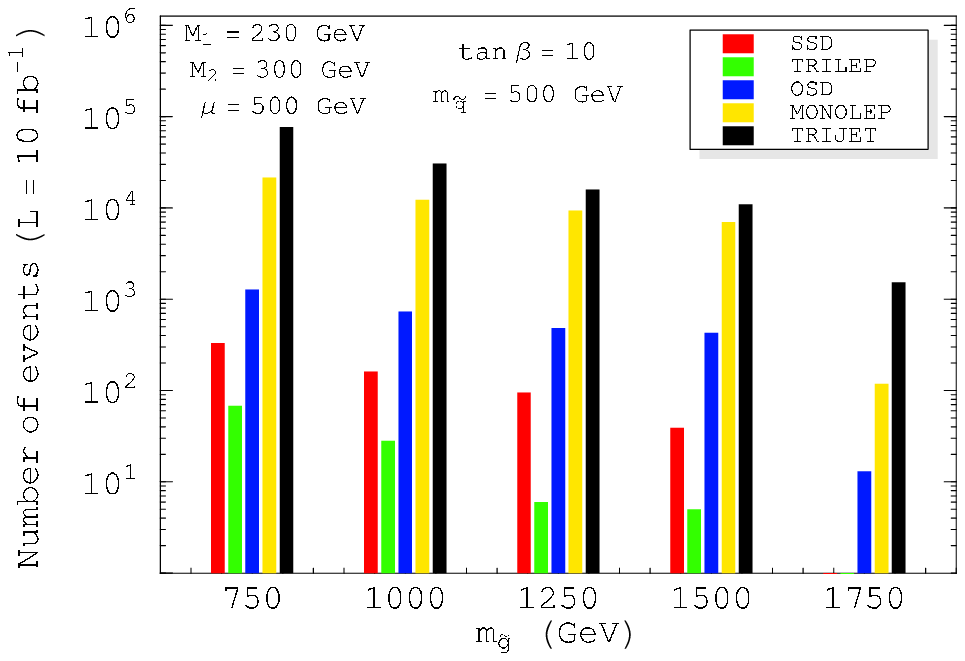,width=9.7cm}
\hspace{-1.7cm}
\epsfig{file=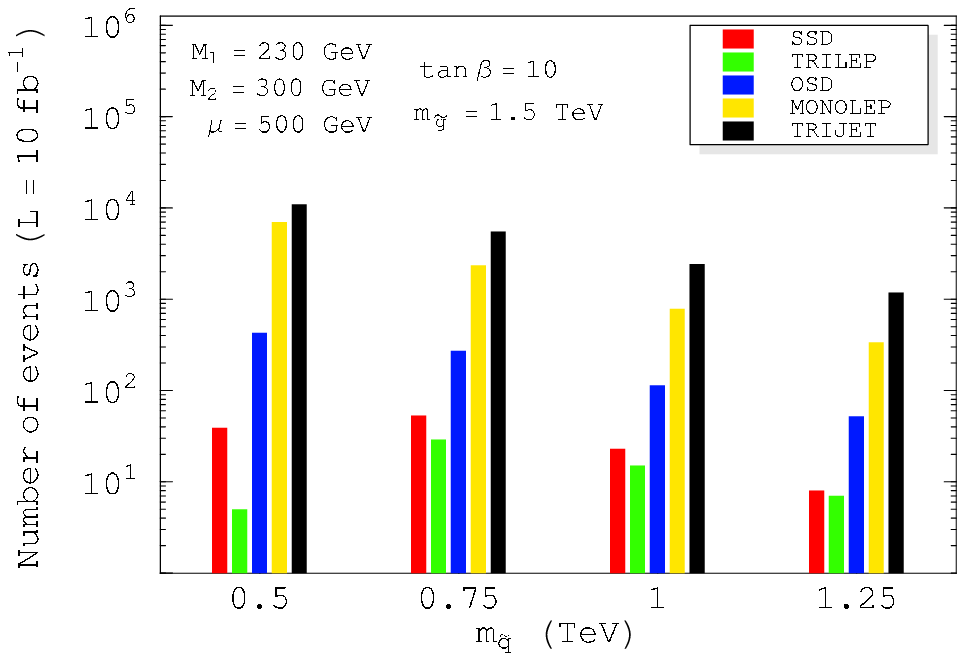,width=9.5cm}}
\caption{Inclusive rates in a somewhat usual paradigm of 
$\mone < \mtwo < \mthree$ with an arbitrary choice of $\mtwo=1.3 \, \mone$
and with (i) $\msquark > \mgluino$ in the upper panel and 
(ii) $\mgluino > \msquark$ in the bottom panel. 
The upper (bottom) left one is with
a fixed squark/sfermion mass of 4 TeV (500 GeV) and varying gluino mass
while for the upper (bottom) right it is the reverse variation with the 
gluino mass fixed at 500 GeV (1.5 TeV). Other relevant parameters are
$\mu=500$ GeV, $\mtwo=300$ GeV, $\tanbeta=10$.}
\end{center}
\vspace*{-1.0cm}
\end{figure}

Table 1 shows how the squarks and the gluinos gradually decouple from
the inclusive signatures with growing masses as is indicated by the
constancy of the numbers. These are residual events coming 
mainly from the electroweak
gaugino production which remains unaffected as squarks and the gluino get
heavier. Table 1 also shows that SSD could be the only signature surviving  
if the gauginos are light enough.

\begin{table}[h!t!b!]
\begin{center}
\begin{tabular}{||c|c||c|c|c|c|c||}
\hline
\hline
     &      &       &            &          &           &         \\
$m_{\tilde{q},\tilde{\ell}}$ & $\mgluino$ & SS &  Trilepton & OS & 1-lepton
& $\geq 3$-\emph{jets} \\
 (GeV)  &  (GeV)  & Dilepton  &            & Dilepton         &           &         \\
\hline
\hline
     &      &       &            &          &           &         \\
1250 & 1250 & 29    &     21     &   122    &    720    &   2462 \\
1250 & 1500 &  8    &      7     &    52    &    337    &   1183  \\
     &      &       &            &          &           &         \\
1500 & 1500 & 10    &      7     &    39    &    219    &    714 \\
1500 & 1750 &  3    &      2     &    18    &    112    &    387  \\
     &      &       &            &          &           &         \\
1750 & 1750 &  3    &      2     &    16    &     87    &    244 \\
1750 & 2000 &  1    &      1     &     9    &     54    &    149  \\
     &      &       &            &          &           &         \\
2000 & 2000 &  1    &      1     &     8    &     43    &    102 \\
2000 & 2500 &  1    &      0     &     5    &     31    &     58  \\
     &      &       &            &          &           &         \\
\hline
\hline
\end{tabular}
\end{center}
\caption{Gradual decoupling of gluino and squarks in the inclusive signatures
via electroweak cascades. All the final states have accompanying 
hard missing transverse energy and the leptonic final states have 
at least two hard jets in addition.
The MSSM parameters are as described in Fig. 1.}
\end{table}

\section{Depleted SSD's and SUSY}
%

If there are regions of MSSM parameter space
where SSD's could be depleted, it would make it initially harder to
identify SUSY.
In the present section, we study some definite but still generic situations
with significantly depleted SSD's and the associated patterns in the MSSM spectrum.

In most studies a
channel is dropped from the discussion once its significance falls below
a critical level of observability. Here we emphasize that a null observation as
part of a pattern with other observations may be very instructive, 
both in identifying the basic new
physics origin (i.e. is it SUSY), and for implications
about the underlying theory.

In Section 2.3 we mentioned some generic situations when
SSD's can be depleted. Here we examine them. \\

%
\noindent
{\bf Case I:} \underline{\bf Heavy charginos} \\

This is the case when either the gluino or the
squark, whichever effectively cascades, is lighter than or almost degenerate 
with the lighter chargino. \\

\noindent
{\bf (a)} \underline{${\mathbf {\mgluino < \msquark}}$:} In this case,
the gluino undergoes the electroweak cascade. When $\mthree < \mtwo$,
the gluino may be lighter than or almost
degenerate with the lighter chargino (including the large radiative 
corrections to the gluino mass).This is an unconventional but possible
region of parameter space.

In the following situations ((a),(b))
we choose masses of gluino and squarks within the reach of the LHC 
(500 GeV and 1 TeV or vice-versa). 
This 
guarantees a large production cross section for these 
sparticles. Later, in situation (c) we demonstrate what happens for larger 
masses with a particular example.

In Fig. 2a, we illustrate how events with SSD's drop out as the lighter chargino
mass approaches the mass of the gluino which is fixed at 500 GeV while the
generic squark mass is 1 TeV.
The variation of 
the chargino mass is implied by that in $\mtwo$ along the x-axis: in
fact we fix $\mu=750$ GeV so the mass of the lighter chargino is close to
$\mtwo$ when $\mtwo << \mu$ as applicable for this figure.
In Fig. 2b, the variation with $\mu$ is shown for a fixed $\mtwo=400$ GeV.
In both cases, the SSD disappear.\\

\begin{figure}[t]
\begin{center}
\centerline{
\hspace{-2.7cm}
\epsfig{file=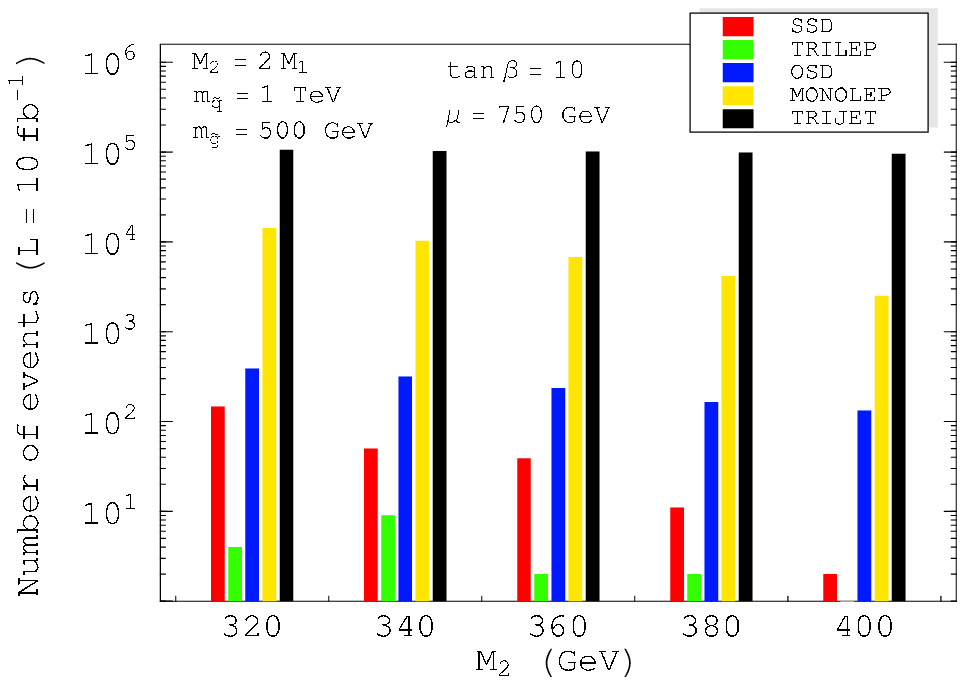,width=9.7cm}
\hspace{-1.7cm}
\epsfig{file=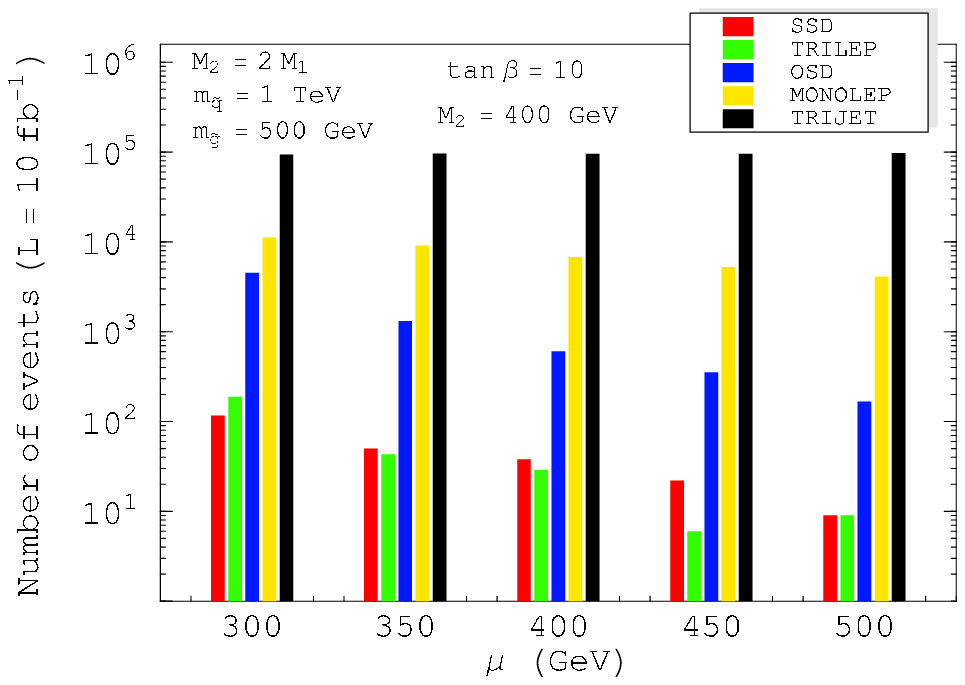,width=9.7cm} }
\caption{Inclusive rates as a function of (a) $\mtwo$ with $\mu=750$ GeV
and (b) $\mu$ for $\mtwo=400$ GeV when 
$\msquark (1 \, {\mathrm {TeV}}) > \mgluino (500 \, {\mathrm {GeV}})$.  
Other relevant inputs are $\mtwo=2 \, \mone$ and $\tanbeta=10$.}
\end{center}
\vspace*{-1.0cm}
\end{figure}

\noindent
{\bf (b)} \underline{${\mathbf{\msquark < \mgluino}}$:} In this case, the 
spectrum is reversed and the squarks undergo the electroweak cascade. 
In Figs. 3a and 3b
we present situations with gluinos as heavy as 1 TeV and squarks 
around 500 GeV when $\mu$ and $\mtwo$ are varied respectively keeping the
other ones fixed.

\begin{figure}[h]
\begin{center}
\centerline{\hspace{-2.7cm}
\epsfig{file=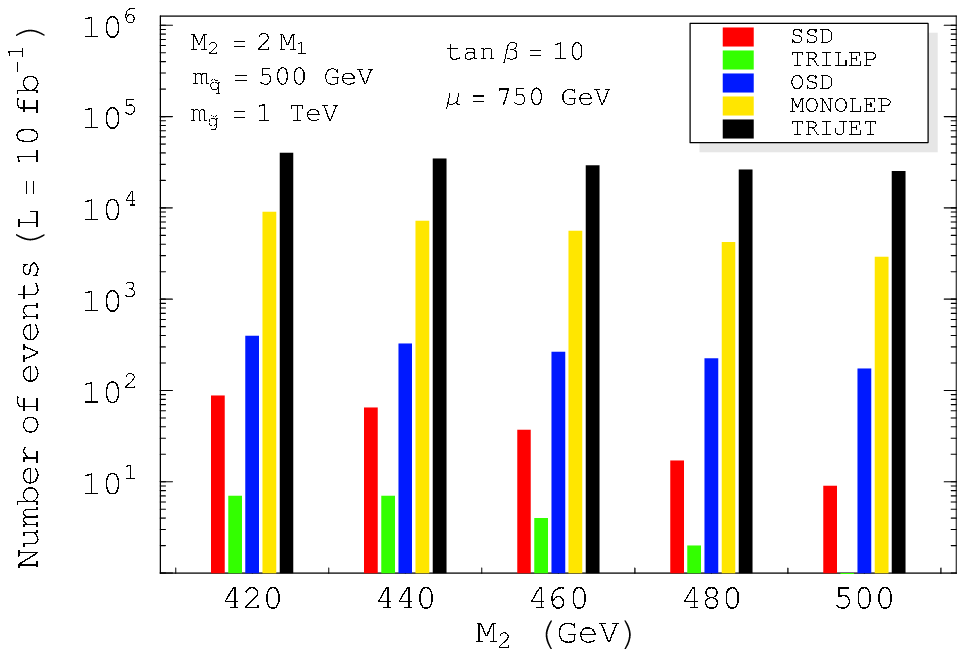,width=9.7cm}
\hspace{-1.7cm}
\epsfig{file=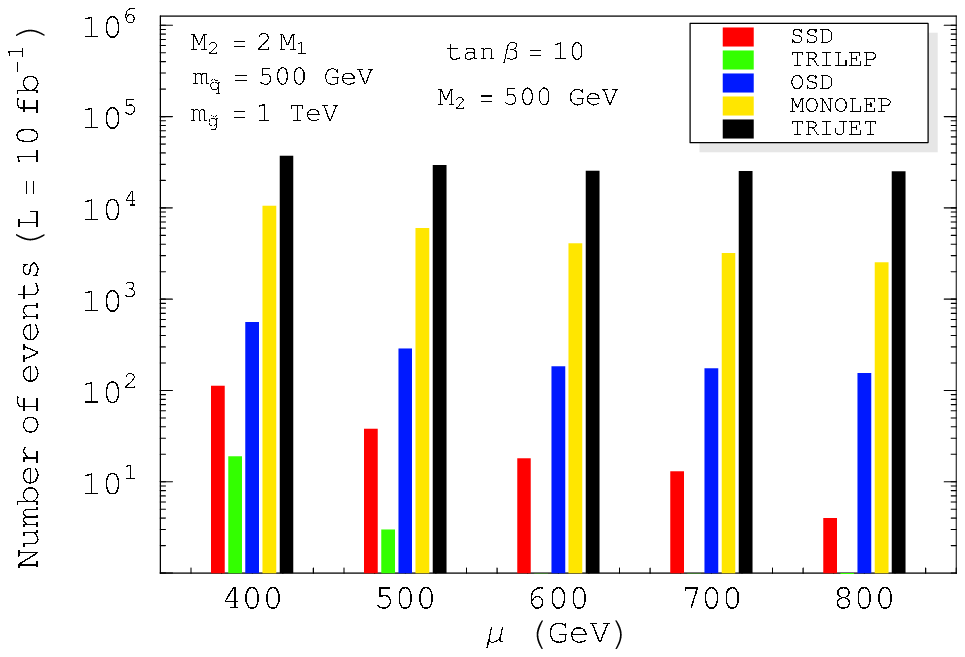,width=9.7cm} }
\caption{Inclusive rates as a function of (a) $\mtwo$ with $\mu=750$ GeV
and (b) $\mu$ for $\mtwo=500$ GeV when 
$\msquark (500 \, {\mathrm {GeV}}) < \mgluino (1 \, {\mathrm {TeV}})$.  
Other relevant inputs are $\mtwo=2 \, \mone$ and $\tanbeta=10$.}
\end{center}
\vspace*{-.50cm}
\end{figure}

\noindent
{\bf (c)} \underline{${\mathbf \msquark \sim 5}$ {\bf TeV} and 
${\mathbf \mgluino=1}$ {\bf TeV:}} Here we discuss a generic situation 
with heavier squarks and gluinos.
For all practical purposes, the squarks cease to contribute to the final 
rates significantly.
The mass of the gluino is still kept relatively low so that it
contributes adequately to the leptonic final states we are interested in.
In Figs. 4a and 4b we illustrate the event rates again as functions of $\mtwo$
and $\mu$ respectively. The event rates drop as the primary
production cross sections decrease. However, the variations with 
$\mtwo$ and $\mu$ still have the same physics explanation as before.

\begin{figure}[h]
\begin{center}
\centerline{\hspace{-2.7cm}
\epsfig{file=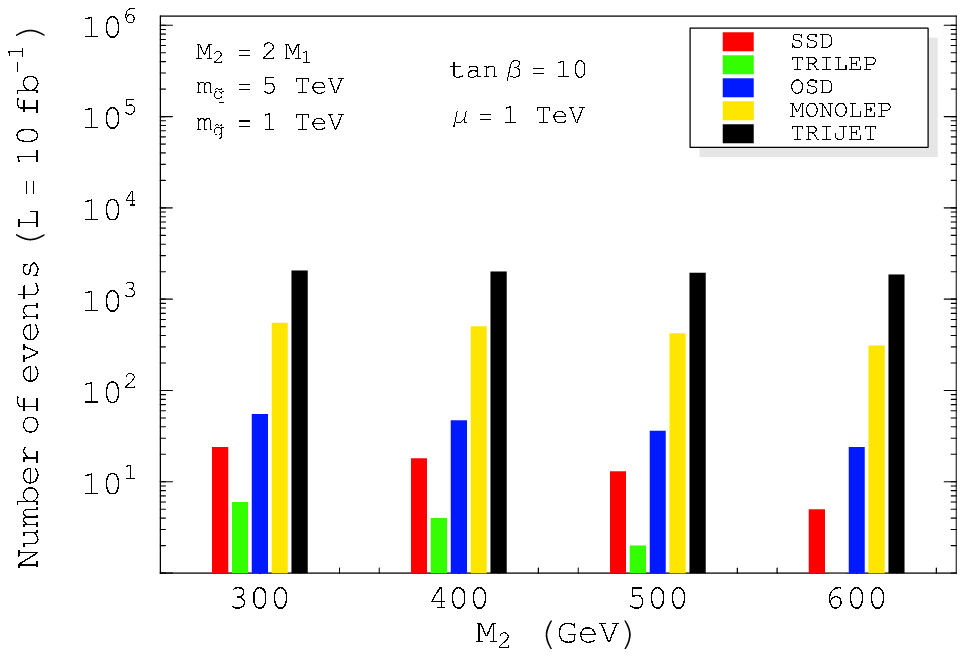,width=9.7cm}
\hspace{-1.7cm}
\epsfig{file=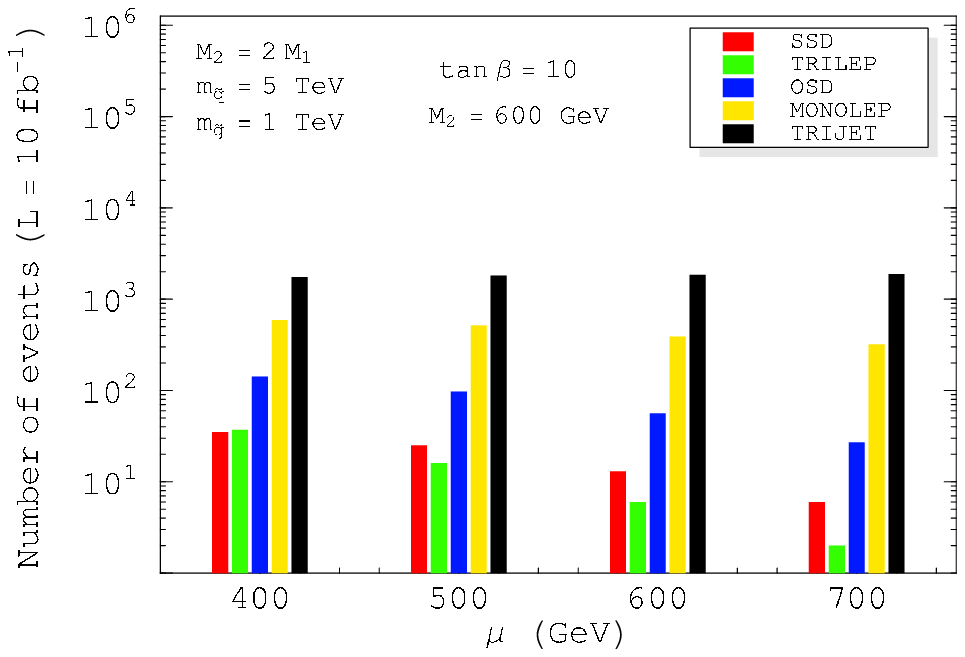,width=9.7cm} }
\caption{Inclusive rates as a function of (a) $\mtwo$ with $\mu=1$ TeV
and (b) $\mu$ for $\mtwo=600$ GeV when 
$\msquark=5$ TeV and $\mgluino=1$ TeV.  
Other relevant inputs are $\mtwo=2 \, \mone$ and $\tanbeta=10$.}
\end{center}
\vspace*{-1.0cm}
\end{figure}

Note that in all these cases the suppression of SSD's occurs 
by suppressing the cascades of gluino/squarks to the lighter chargino. 
On the other hand, even for such parameters the second lightest neutralino always turns out to be 
roughly degenerate with the lighter chargino and similar in composition, given the structure of
the chargino and the neutralino mass matrices in the MSSM. 
Inevitably, electroweak cascades of squarks/gluino 
to the second lightest neutralino, which
is another source of leptons under cascades, get suppressed.
Thus, in general, depletion in SSD's
is not an isolated phenomenon but is accompanied by depletion in other leptonic
channels. However, given different combinatoric factors involved
in determining the rates of different leptonic final states they get 
affected in somewhat different proportions, as is seen from Figs. 2-4.  

The SSD's and the trileptons have a similar rate.
In Figs. 2-4 we see that the trileptons lose out to the SSD's and deplete
to unobservable levels before SSD's do. The channels surviving when both of these
get critically depleted are the OSDL+jets, 1-lepton+jets and the inclusive
jet events.
While the depletion gives a pretty good idea of the relative 
hierarchy of the EW gaugino masses, the event rates for the surviving modes 
(especially the jets) hint at the squark and 
gluino mass scales and their relative hierarchies. When combined with clues
from the electroweak gaugino sector this could shed light on 
non-universality of gaugino masses.

For Figs. 2-4, we maintained to the relation $\mtwo = 2 \mone$ for the $SU(2)$ and
the $U(1)$ gaugino masses. As is well known, this is what happens in 
mSUGRA type scenarios with
gaugino mass unification at a high scale. In Fig. 1 we also considered a somewhat
different relation $\mtwo = 1.3 \mone$ as an alternative to
unification. Now we study
the impact of gaugino mass relations on the leptonic events. 

The relation between $\mone$ and $\mtwo$ affects the leptonic rates
by altering the branching fractions (BF) of gluinos or squarks 
to the lighter chargino and the second lightest neutralino.  The latter two
are two important sources of leptons in cascades at hadron colliders.
We fix $\mtwo$ so that the ratio of $\mthree/\mtwo$ remains fixed
while varying $\mone$. With increasing (decreasing) $\mtwo/\mone$ the 
mass of the LSP decreases (increases) which in turn affects the decays
of squarks and gluinos to the LSP, the lighter chargino and the second
lightest neutralino. For $\mtwo/\mone \gtrsim 1$ the BFs compete and the BF
to lighter charginos usually dominates over that to a rather heavy LSP.
Under such circumstances, one still expects copious production of SSD's.
However, as the LSP mass approaches the mass of the lighter chargino,
the decay $\chpm1 \to \ntrl1 \ellpm \nu$ gets suppressed thus depleting
the leptons and hence the SSD's. On the other hand, with
growing $\mtwo/\mone$ the LSP mass decreases and the BF to LSP increases 
thus depleting
the leptons (and SSD's). This does not continue long as the competing BFs
get saturated, and does not deplete SSD's beyond a certain
point. For $\mtwo/\mone=2$ such a saturation is already reached.

\vskip 10pt
\begin{table}[h!t!b!]
\begin{center}
\begin{tabular}{||c||c|||}
\hline
\hline
$\mtwo/\mone$ & SSD-events \\
\hline
3.00 & 29 \\
2.00 & 36 \\
1.50 & 94 \\
1.25 & 133 \\
1.10 & 117 \\
1.05 & 48 \\
1.00 & 4 \\
\hline
\hline
\end{tabular}
\end{center}
\caption{Variation in the rate of Same-Sign Dileptons as a function of
$\mtwo/\mone$ with $\mtwo=350$ GeV. The other relevant parameters are the
same as in Fig. 2a, i.e., $\mgluino = 500$ GeV, $\msquark=1$ TeV, 
$\mu=750$ GeV and $\tanbeta=10$.}
\end{table}

In Table 2 we present the variation in the rates as a function of 
$\mtwo/\mone$. We chose the inputs from 
Fig. 2a for direct comparison and fix $\mtwo=400$ GeV. The trends are 
clearly as expected and as discussed above. 

In most of the cases in Figs. 2-4 the trilepton final state gets
a similar or worse suppression than the SSD's. But the OSDL or 1-lepton final
states could still be statistically significant thanks either to the 
combinatorics or the
low lepton-multiplicity involved or both. Such a situation suggests a
near degeneracy of the
squarks/gluino and the lighter electroweak gauginos. However, one 
should keep in mind that unlike SSD's these leptonic channels are likely to have
significant SM backgrounds which might not be put under control with the 
naive ATLAS-type
cuts \cite{atlas} 
we employ here (which have mainly been studied for mSUGRA scenarios
and need further examination). 


\vskip 20pt
\noindent
{\bf Case II:} \underline{\bf When $\mathbf{\msquarkr < \mgluino < \msquarkl}$}
{\bf or} \underline{$\mathbf{\mstop1 < \mgluino < \msquarkl \simeq \msquarkr}$} \\

\begin{figure}[h]
\begin{center}
\centerline{\hspace{-2.7cm}
\epsfig{file=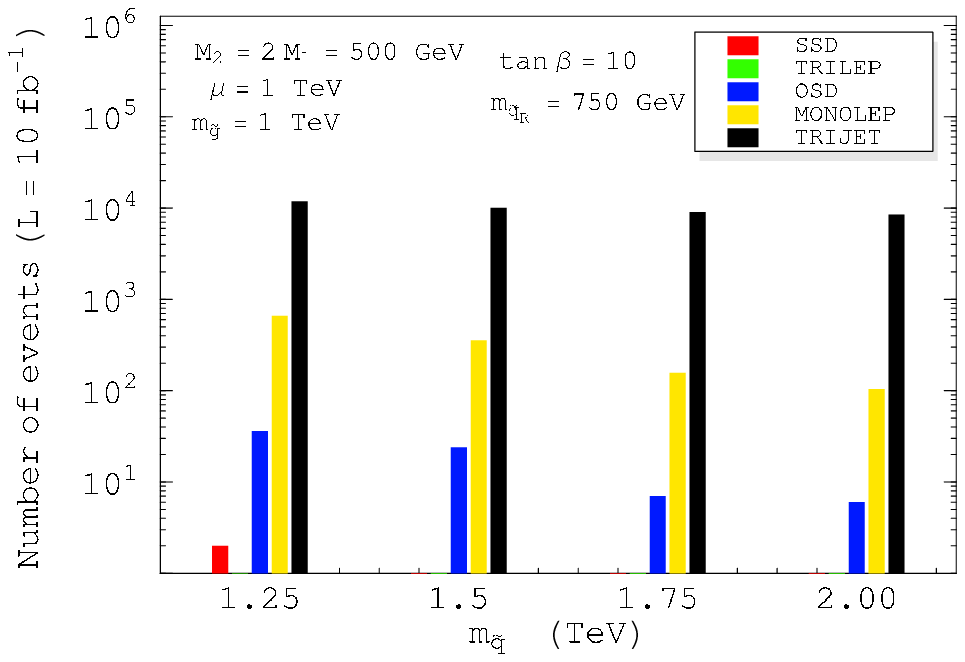,height=5.3cm}
\hspace{-1.7cm}
\epsfig{file=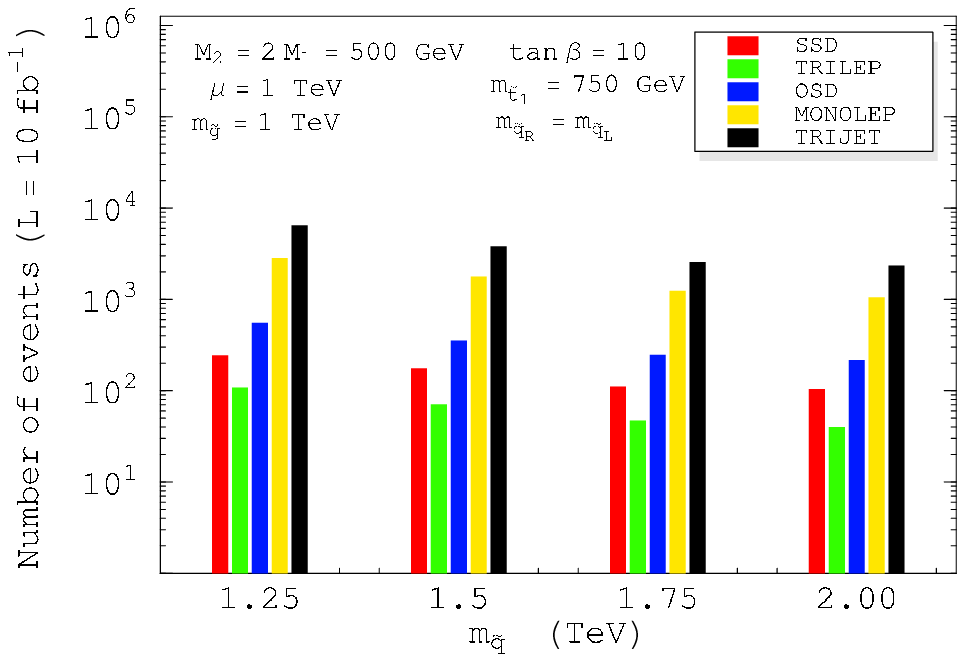,height=5.3cm} }
\caption{Variation of inclusive rates as a function of $\msquarkl$ for
$\msquarkr < \mgluino < \msquarkl$ (left) and 
$\mstop1 < \mgluino < \msquarkl=\msquarkr$ (right), for
$\mgluino=1$ TeV and $\msquarkr (\mstop1)=750$ 
GeV. Other relevant SUSY paramters are $\mtwo=2\mone=500$ GeV, 
$\mu=1$ TeV, $\tan\beta=10$.}
\label{light-squarkR}
\end{center}
\vspace*{-1.0cm}
\end{figure}

\noindent
A spectrum like this can arise from non-universal soft scalar masses at a high
scale with natural splitting between left and right handed ones 
\cite{Kawamura:1994ys, Kawamura:1993uf, Datta:1998uv, Datta:1999uh}. 
Also, $\mstop1$ can be comparatively light at the weak scale even in 
universal scenarios as $m_{\tilde{t}}$ runs down faster due to its large yukawa 
coupling.
Apart from that, from a pure MSSM standpoint
this kind of a spectrum could have interesting phenomenology.
Here, all left handed (or heavy) squarks would decay mostly to gluino
and a quark followed by the gluino decaying to right handed squark 
(or $\sstop1$) and a quark (top quark).
The right handed squark mostly undergoes a decay to quark and the LSP thus
depleting leptons in the final states. The phenomenon is illustrated
in the left side of Fig. \ref{light-squarkR}. 

If $\sstop1$ is the only squark 
lighter than the gluino, the result is a little bit involved. 
The gluino would have a 100\% (in contrast to only a few percent 
when all other right handed 
squarks are lighter than the gluino) 
decay branching fraction to $t \sstop1$ shared
equally by Br[$\gluino \to \sstop1 \bar{t}$] and Br[$\gluino \to \sstop1^* t$].
Note this ensures having some same sign tops from the decay of gluinos which 
in turn
would lead to the SSD. In fact, the associated stops could boost the 
SSD count through combinatorics if Br[$\sstop1 \to t \ntrli$] dominates. 
This gives leptons from top quarks.
Hence, enhancements in all the leptonic channels are present and illustrated
in the right of Fig. \ref{light-squarkR} for a similar variation of the
heavier squark mass scale as for the left figure. 
Final states with 4 tops is a very likely possibility. 

Left handed squarks may still have significant branching fractions to quarks 
and electroweak gauginos if they are not heavy compared to the gluino. 
Decays of these gauginos may result in some leptons.

However, as expected, and can be seen in Fig. \ref{light-squarkR}, with
increasing masses for the left handed squarks (with the masses of the gluino
and the right handed squarks kept fixed) the strong decay branching fraction
for $\squarkl \to q \gluino$ starts dominating and depleting the
leptons in the final state. 

In any case, the SS dileptons and the 
trileptons get heavily depleted to an unobservable level for such a spectrum.

\vskip 20pt
\noindent
{\bf Case III:} \underline{\bf When the radiative 
decay ${\mathbf \gluino \to g \ntrli}$ is significant} \\

As pointed out in Section 2.3, the spectrum here is reminiscent of a Split 
SUSY scenario \cite{Arkani-Hamed:2004fb, Giudice:2004tc, Arkani-Hamed:2004yi} 
with very high squark masses 
($ \ge {\cal{O}}(10)$ TeV). A gluino, which is comparatively light 
becomes the only
interesting strongly interacting sparticle to study. Decays of gluinos in such a
framework have recently been studied in Refs.\cite{Toharia:2005gm,
Gambino:2005eh}. 
In the first reference, an interesting scenario was discussed, in
which stops are lighter than the other squarks. Depending on the
gaugino-higgsino spectrum, 
the 2-body radiative decays can then
dominate by the interplay of various suppressions in the 3-body decays. 
Radiative decays are enhanced inside the loop by lighter stop
masses, and they also benefit from large logs (this is a Split SUSY effect,
requiring the presence of some higgsino content in the
neutralino). There is also some phase space enhancement of 2-body
versus 3-body decays.
Finally the dominant 3-body decays of the gluino into a final state
containing at least one top (because of lighter stops) might be
kinematically forbidden or suppressed (given the large mass of the top
quark this is not an unreasonable situation). 
In this case, 2 gluon jets + missing energy might be the main
signal for SUSY at LHC, and leptons would be completely depleted in the signal.
It is something to keep in mind if signals for physics BSM prove 
themselves hard to find at LHC.

We did not find any generic region of parameter space where the inclusion of
radiative decays of the
gluino\footnote{We do not know of any event 
generator which includes these radiative gluino decay modes. The only
publicly distributed decay code known to us that includes these modes
is SDECAY \cite{maggie}. Since Pythia uses its own decay routines, we had to
supplement it with the results from SDECAY.}
could affect an otherwise healthy rate for SSD significantly 
for not too heavy a squark mass
($\lesssim 5$ TeV). The reason is simple. To deplete the SSD we need a handful 
of them from the decays of charginos which in turn demands
a significant Br[$\gluino \to \chpm1 q \qbarprime$] to start with. Note that
the latter requires a high gaugino content in the chargino (unless $\sstop1$
mediated gluino decay becomes important) and ensuring that
inevitably increases the gaugino content of the LSP. As pointed out in 
Ref.\cite{Toharia:2005gm} this does not favor BR[$\gluino \to \ntrl1$]. Hence,
a mixed gaugino-higgsino region of the SUSY parameters would be
needed. 
But such an arrangement requires the LSP to be heavy 
enough ($\sim \mchpm1 \lesssim \mgluino $) such that 
$\gluino \to q \qbar \ntrl1$ could gradually start
losing out to the radiative 2-body decay. In the absence of heavy 
squark propagators (as in Split SUSY) this is the only way to prevent
$\gluino \to q \qbar \ntrl1$ from leading. This leads to
a rather fine-tuned (but possible) region of parameter space by requiring at the same time
not too degenerate a $\chpm1$ and a $\ntrl1$ such that the leptons from the 
chargino decay are still observable.


\vskip 10pt
\noindent
{\bf Case IV:} \underline{\bf When lighter chargino is accessible down the cascade
but}\\ \hspace*{0.65in} \underline{\bf degenerate with the LSP} \\

\vspace{-.5cm}

\begin{center}
\begin{figure}[h]
\vspace{0cm}
\hspace{-2.7cm}
\includegraphics[width=9.75cm]{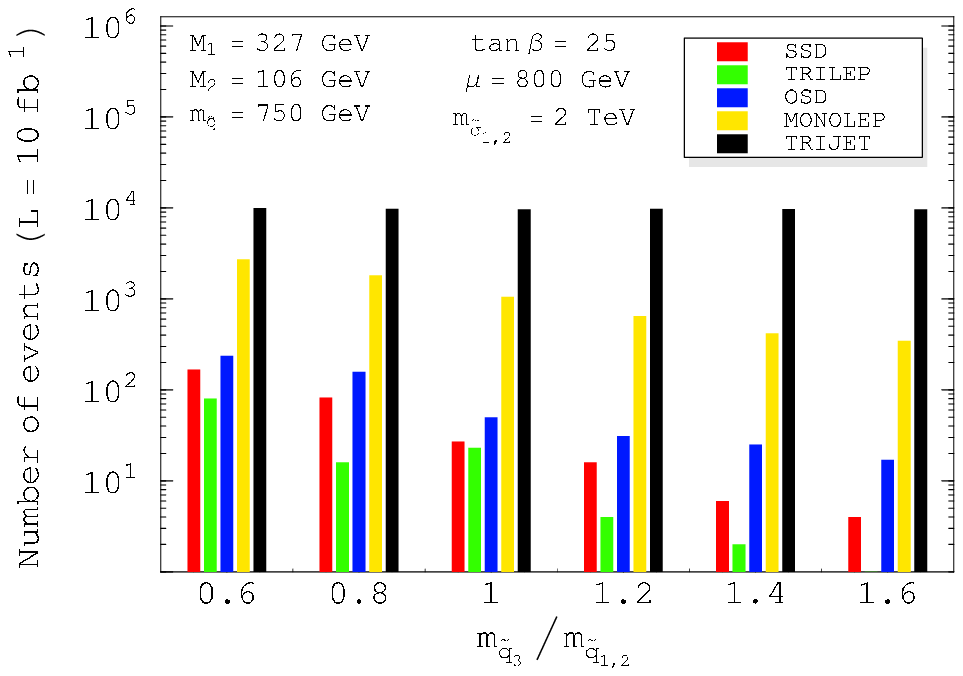}
\hspace{-1.9cm}
\includegraphics[width=9.75cm]{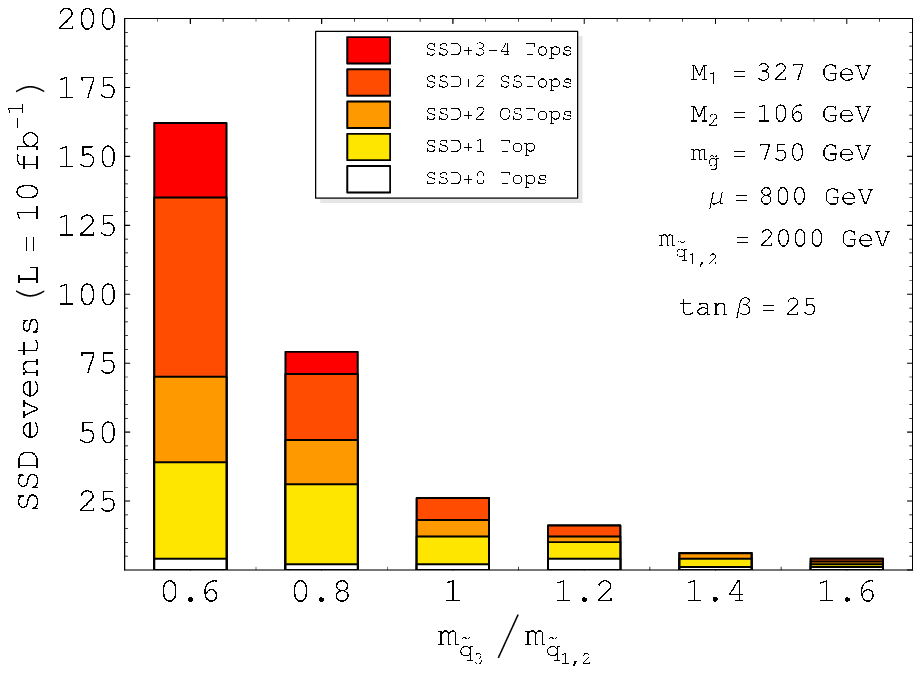}
\vspace{-0.8cm}
\caption{Event-rates for different lepton and jet final states 
(left) and top quark content in SSD events (right panel)
as a function of the ratio between third and lighter generation 
squark masses. Shown in the figures are the MSSM inputs used: a
typical AMSB gaugino spectrum with $\mu$ chosen to be larger than the
mass of the gluino, and with $\tan\beta=25$ (we checked that with a value
of $\tan\beta=5$ we did obtain results with no visible differences).
The heights of the respective bars on the right figure
directly correspond to the total SSD event-rates in the left figure and
the length of each shaded portion on the right indicates the 
proportional contributions.}
\label{varyrsq3}
\end{figure}
\end{center}
\vspace{-.25cm}

\noindent
In this case, leptons from the decay
$\chpm1 \stackrel{\tilde{\ell}^*, \tilde{\nu}^*} \longrightarrow
\ell \nu \ntrl1$ would be too soft to be observed because of the
chargino-LSP degeneracy. This depletes leptons in the final 
states and SSD in particular.
The presence of soft charged pions and a displaced vertex from the decay 
$\chpm1 \to \pi^\pm \ntrl1$ would be very characteristic in this
situation \cite{Gherghetta:1999sw}.

We consider such a spectrum as a part of the MSSM. 
As already noted, the issue central to 
depleting leptons is the mass-hierarchy among the gauginos 
(including the gluino) and the squarks. Hence, we freely vary 
these soft parameters as before within the framework of the MSSM. 
Thus, we end up with a spectrum which looks like that of the AMSB 
gaugino sector (but without restricting ourselves to that specific
scenario); i.e. the LSP is Wino like and almost degenerate 
with the lighter chargino. The second lightest neutralino is
Bino-like. The gluino is heavier than these electroweak gauginos.
The squarks, the sleptons and the higgsinos are all more massive.
Thus, production of gluinos and their subsequent decays constitute
the dominant SUSY process at the LHC. In such a scenario we focus on
the hard \emph{lepton+jets}$+ \not \!\!E_T$ events.

In such a scenario
SSD events are generally associated with multiple top quarks.
The reason for this is as follows. 
Decays of the lighter chargino would not lead to SSD events
for the reason mentioned above. 
Also, the second lightest neutralino being Bino-like 
(for heavy enough $\mu$) its decay to $W^\pm$ and Wino-like 
lighter chargino, $\ntrl2 \to \chpm1 + W^\mp$  
is heavily suppressed. 
This rules out another source of leptons (and specifically
SSD) in $W^\pm$'s.
With heavier gauginos somewhat decoupled 
(with masses $\cal{O}(\mu)$), the
only significant source for SSD's are $W$'s coming from the decays 
of Same-Sign top quarks, which in turn appear
in decays of the gluino into charginos and/or neutralinos:
\begin{eqnarray*}
\gluino \stackrel{\tilde{t}^*,\tilde{b}^*} \longrightarrow  
t t(b) \ntrli (\chipm)  \quad \mathrm{with} \quad t \to b W^\pm .
\end{eqnarray*}
Thus charges on hard leptons could be traced back to those
on the top quarks pointing to Same-Sign or 
Opposite-Sign top quark pairs down the cascade. 
Of course, Same-Sign top pairs or bottom pairs could be as valuable a
signature as SSD in general.

\begin{figure}[t]
\vspace{0cm}
\center
\hspace{-3cm}
\includegraphics[width=12cm]{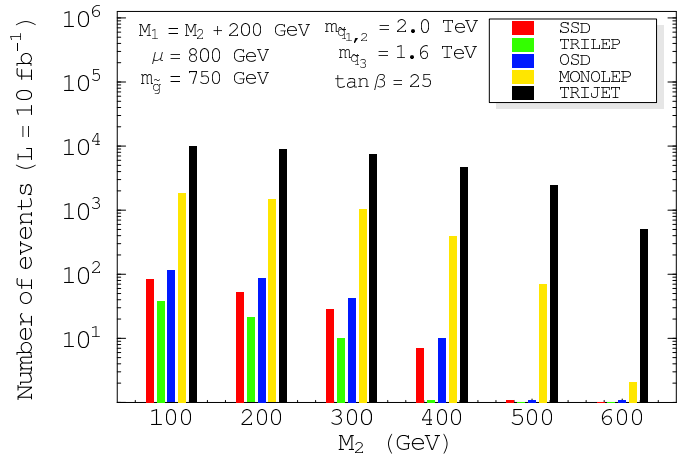}
\vspace{-.3cm}
\caption{Variation in different event rates as a function of $\mtwo$,
such that $M_1=M_2+200$ GeV, $\mgluino=750$ GeV, and $\mu$ is fixed at
$800$ GeV. The mass of third generation squarks
is $m_{\widetilde{q}_3}=1.6$ TeV while for the first and second generations
we take $m_{\tilde{q}_{1,2}}=2$ TeV. As the gaugino masses $M_1$ and
$M_2$ increase while maintaining the gluino mass fixed, the phase space
for top events in gluino decays is reduced and so are the SSD events, which
in this scenario all come from top decays.} 
\label{varym2}
\end{figure}

One would therefore expect the hard lepton signals to gradually 
turn weaker as stops and/or sbottoms grow in mass
compared to squarks from the first two generations. 
In fact, this is reflected in the left of Fig. \ref{varyrsq3}.
When varying the value of the third generation squarks mass up to 
1.6 times the generic squark mass for the lighter generations,
the SSD, the OSD and the trilepton 
signals weaken to an unobservable level for a 10 fb$^{-1}$ of LHC
data.

The 1-lepton final state, though weakening, continues to enjoy 
the usual combinatorial advantage of low-multiplicity 
and still has a healthy rate. Since hard jets predominantly come 
from initial cascades, the 3-jet inclusive rate, in our case, 
gets a predominant contribution from jets in the 
$\gluino$ decay. 
Squarks with masses around 2 TeV are sub-dominant at their
production level, more so for 
individual squark flavors like $\tilde{t}$ or $\tilde{b}$. Hence,
increasing these masses does not affect the jet rates noticeably.

In the right panel we show the admixture of top-multiplicities and
their charge-content in
the SSD events as a function of the same variable as for the 
left figure.
Clearly, `topless' SSD events are extremely rare while the majority
of SSD events contain two top quarks. 
Such an association of SSD events with top quarks seems to be very
characteristic of a AMSB-like gaugino mass hierarchy
(when $\mu > \mgluino$ such that decays like 
$\gluino \to \chi_{_{3,4}}^0, \chi_{_2}^\pm$ are kinematically
forbidden) and could help understand possible new events at the LHC.

In Fig. \ref{varym2} we show that the hard lepton signals would 
also get reduced to an unobservable level with increasing Wino and 
Bino masses, because the phase space for top events gets reduced in the decay
$\gluino \stackrel{\tilde{t}^*,\tilde{b}^*} \longrightarrow  
t t(b) \ntrli (\chipm)$.
Significantly enough, even the single lepton events could disappear 
for $M_2$ as heavy as 600 GeV in our present example.
Although reduced, the trijet channel remains still quite populated
for the same reasons discussed before.

\vskip 20pt
\noindent
{\bf Case V: \underline{The Higgsino LSP region}}
\vskip10pt

\begin{center}
\begin{figure}[h]
\centerline{
\hspace{-2.7cm}
\epsfig{file=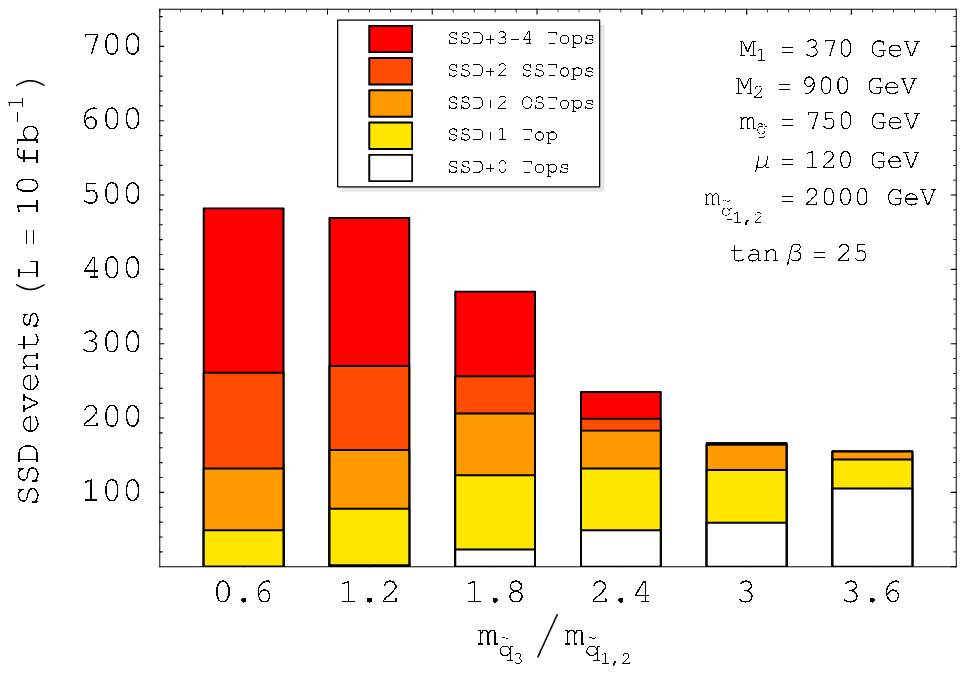,
width=9.75cm}
\hspace{-2cm}
\epsfig{file=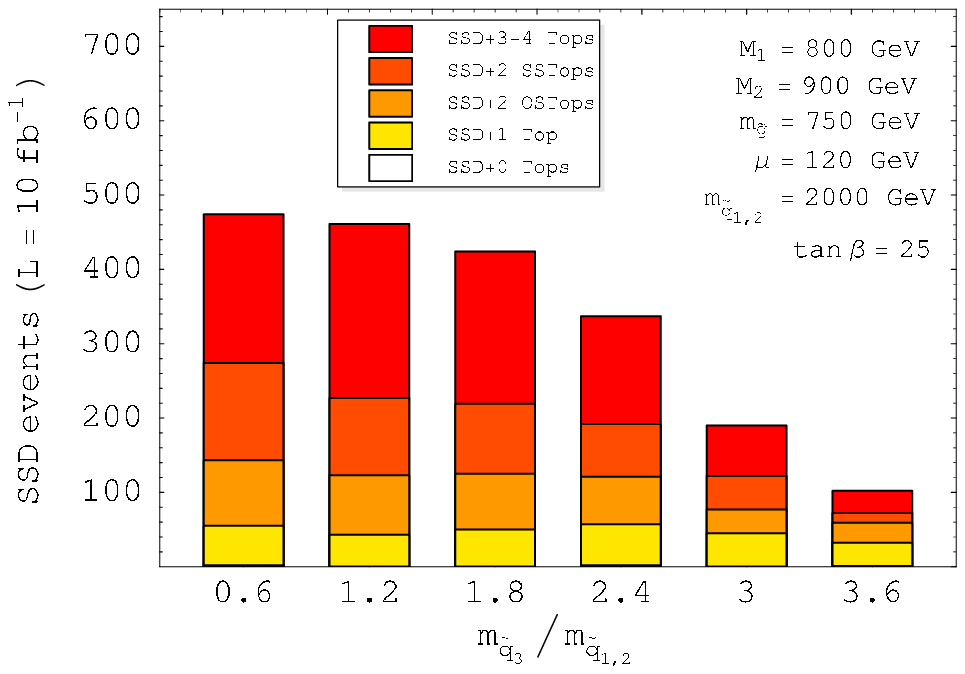,
width=9.75cm} }
\caption{Same sign dilepton rate in the higgsino LSP region and the
multiplicity of the associated top quarks for
(a) $\mu << \mone < \mgluino$ (left) and 
(b) $\mu < \mgluino < \mone$ (right). The input parameters used are
indicated in the figures.}
\label{higgsinolspvaryrsq3}
\end{figure}
\end{center}

\noindent
Next we look into the case where the gluino decays to 
lighter chargino which is almost a pure higgsino, unlike in
Case II where it was mainly Wino. This is achieved by having
$\mu < \mone,\mtwo$.
In this case $\ntrl1, \ntrl2$ and $\chpm1$ constitute
the set of lighter gauginos (and also the lightest of the sparticles)
with $\mntrl1 \sim \mntrl2 \sim \mchpm1 \sim \mu$ 
and all of them are higgsino dominated. The third lightest neutralino
used to have a mass $\mntrl3 \sim min(\mone,\mtwo)$ 
and be correspondingly Bino- or Wino-like while the reverse is true for 
the heaviest neutralino and the heavier chargino.
Here, for the purpose of demonstration, we will consider two cases: 
(a) $\mu << \mone < \mgluino$ and (b) $\mu < \mgluino < \mone$.  
In both cases we will forbid gluino-to-Wino decays by fixing 
$\mtwo$ large ($\mtwo=900$ GeV).

For case (a), the main phenomenological difference with the 
AMSB-like scenario is that the Binos ($\ntrl3$ in this case) 
produced from 
the gluino decays do couple with the higgsino-like lighter chargino and
thus could undergo the decay 
$\ntrl3 (\tilde{B}) \to \ellpm \nu \chpm1 (\tilde{H}^\pm)$. These 
leptons can be hard enough for large $\Delta M = \mone - \mu$.
Thus we get an extra source of leptons in the form of 
$\ntrl3 (\tilde{B})$ which, unlike the 
AMSB-like case, does not necessarily\footnote{Of course, 
$\gluino \to t \bar{t} \chi_{1,2,3}^0 \; \mathrm{and} \;
\gluino \to t b \chpm1$ are all  possibilities which would involve
top quarks. However, unlike previously, this time these top quarks 
are not unique sources of observable leptons.} 
have accompanying top quark(s).
Indeed, the left of Fig. \ref{higgsinolspvaryrsq3} 
shows how topless SSD events first gain significance and then 
dominate as stops and sbottoms grow in mass.

In case (b), when gluinos only have access to the light
higgsinos we expect a similar behavior as in AMSB, namely a direct 
association of SSD and top events. The right panel of 
Fig. \ref{higgsinolspvaryrsq3}
illustrates this correlation. However, this time, a comparison with
the right panel of Fig. \ref{varyrsq3} reveals that 
a depletion in SSD events is harder to achieve with an increasing 
ratio of
third to lighter generation squark masses, because of 
the yukawa enhanced couplings between (s)tops and higgsinos
(unlike (s)tops and the winos in the AMSB-like scenario).

\begin{figure}[h]
\begin{center}
\hspace{-3cm}
\centerline{\epsfig{file=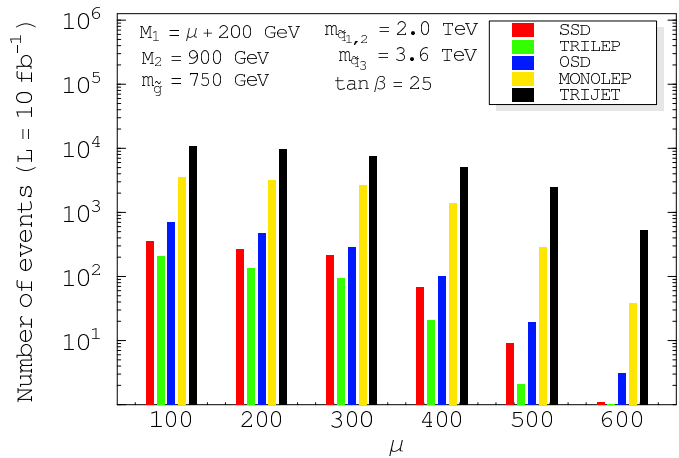,width=12cm}}
\vspace*{-.5cm}
\caption{Depleting leptons by increasing $\mu$ and $\mone$ 
simultaneously. The mechanism of depleting leptons in the higgsino 
LSP region has in it a parallel of what shown in Fig. \ref{varym2}
with varying $\mtwo$.}
\label{higgsinolspvarymu}
\end{center}
\vspace*{-.5cm}
\end{figure}

Figure \ref{higgsinolspvarymu} shows that by increasing both 
$\mu$ and $M_1$ starting at the values of case (IIIa), in which gluinos 
can decay both into Binos and Higgsinos, we only manage to
reduce the hard lepton signals to minimal levels
 when $M_1 >\mgluino$
and when tops become
kinematically forbidden in gluino decays (with $\mu=600$ GeV and
$M_1=800$ GeV).
Of course we can also eliminate the 
leptons in case (IIIb) above by increasing $\mu$ (\emph{i.e.} by 
making the lighter higgsinos heavier, as shown in Fig. 
\ref{higgsinolspvarymu}). With the Bino channel closed from the
start, top quarks gradually 
get kinematically disfavored in the decay 
$\gluino \to t \bar{t}(\bar{b}) \ntrl1(\chpm1)$ in a way 
similar to when $\mone$ and $\mtwo$ were increased
to achieve the same effect as illustrated in Fig. 6.

Then, just as in the AMSB case, SSD events do exist but do not pass
the usual cuts due to their extreme softness.

\vskip 20pt
\noindent
{\bf Case VI:}  \underline{\bf When $\mathbf{\mntrl1 < \msnu < \mntrl2 \simeq \mchpm1} < {\mslep}_{_L}$} \\

\noindent
This is the scenario with ``effective'' (or ``virtual'') 
Lightest SUSY Particles (ELSP or VLSP) 
\cite{Datta:1994un, Datta:1994ac, Datta:1996ur}.
The cascade $\ntrl2 \to \nu \snu \to \nu \ntrl1$ has a 100\% branching
fraction. This ensures both $\ntrl2$ and $\snu$ decay invisibly, ``effectively''
like the LSP. 
$\ntrl2$ ceases to be a source
of leptons thus depleting them in the final state.

For the lighter chargino, which is almost
degenerate with $\ntrl2$, 
Br[$\chpm1 \to \ellpm \snu \to \nu \ntrl1 \ellpm$] 
would be 100\%. For small $\Delta m = \mchpm1 - \msnu$
(which is the case as we will see below)\footnote{Such a spectrum can easily 
be accommodated in popular SUSY-GUT scenarios with gaugino mass 
unification at the GUT scale\cite{Datta:1998yw}.} the leptons
could be too soft to trigger on, so $\chpm1$ would decay
invisibly \cite{Datta:1998yw}.

The mass-splitting between $\snu$ and its partner
slepton in the doublet (at the lowest order) is given by
\begin{eqnarray*}
m_{{\ell}_L}^2 = m_{\snu}^2 - m_W^2 \cos 2\beta 
\label{su2split}
\end{eqnarray*}
and constrained by the $SU(2)$-breaking $D$-term.
The splitting is thus a function of $\tan\beta$ and goes down
sharply with increasing slepton mass.
For example, the achievable splitting 
for $\mslep \approx 500$ GeV
(1 TeV) is around 7(3) GeV for large $\tan\beta$.

\begin{figure}[h]
\begin{center}
\vspace*{.2cm}
\centerline{
\hspace{-3cm}
\epsfig{file=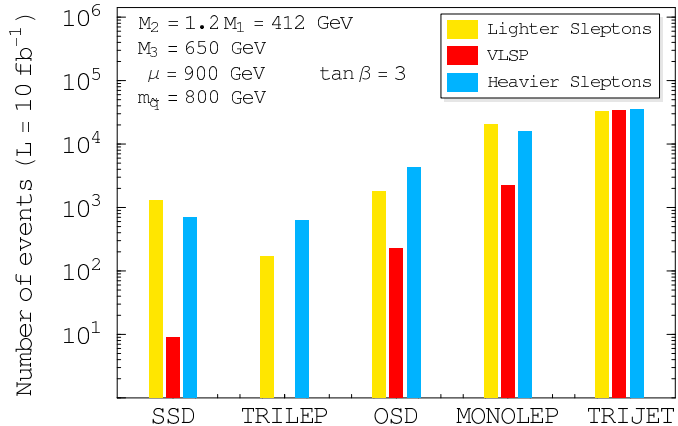,width=12cm}}
\vspace*{-.5cm}
\caption{Leptonic activities for a VLSP spectrum and spectra just
away from it. The VLSP spectrum in MSSM 
is generated through Suspect v2.33 with
the following major inputs: $\mtwo=1.2 \mone=412$ GeV,
$\mslepl=\mslepr=412$ GeV except for $m_{\tilde{\tau}_R}=419$ GeV,
$\mthree=650$ GeV, $\msquarkl=\msquarkr=800$ GeV,
$\mu=900$ GeV and $\tan\beta=3$. For heavier slepton case
the slepton masses were set at 425 GeV while for the lighter slepton
analysis a value of 400 GeV was set. Rest of the inputs were kept 
fixed to those in the VLSP case.}
\label{vlsp_variance}
\end{center}
\vspace*{-.5cm}
\end{figure}

In Fig. \ref{vlsp_variance} 
we present the leptonic activity for a VLSP spectrum (in red, 
the middle bars) and compare it to other situations.
For the choice of inputs, see the figure and its caption.

A very low SSD count for an
ideal VLSP spectrum is not due to a suppressed leptonic branching 
fraction\footnote{
Br[$\chpm1 \to \snu \ellpm$]=100 \% which could potentially be a copious source
of SSD with gluinos effectively cascading.} of $\chpm1$, but this is an
artifact of a small $\Delta m$ when the
emerging lepton is too soft to pass the $p_T^{lepton}$ trigger. 

As one makes the sleptons
lighter two things happen. Charged sleptons could become kinematically
accessible in the 2-body decay of $\chpm1$. 
The leptons in the decay $\sleppm \to \ellpm \ntrl1$ 
could easily be hard enough 
(when $\Delta m^\prime=\mslep - \mntrl1$ is large 
enough, which is the case here) to pass the $p_T^{lepton}$ cuts and thus
contribute to the SSD signal. For even smaller slepton (and hence sneutrino)
masses, leptons from $\chpm1 \to \snu \ellpm$ start getting harder 
as well thus reinforcing the SSD count. This is well described
by the green bars (the leftmost bars).

For heavier sleptons, the 2-body decay modes of the lighter gauginos close up.
Democracy is thus restored among all possible 3-body decay modes invoving
both leptons and the jets with usual dependence on the masses and the
couplings.  These tend to restore the lepton counts in the final state. 
In Fig. \ref{vlsp_variance}, 
the effect is presented in blue bars (the rightmost ones).

The discussion again holds for the trilepton counts.  However, for OSDL and
single leptons channels the combinatorics and/or lepton-multiplicity ensure
already healthy rates even for the VLSP spectrum. In these cases, 
slipping to either side of the spectrum leads to nothing drastic.

On the other hand, the inclusive 3-jet signal seems insensitive to such 
variations of spectrum.
This indicates that these jets are predominatly primary jets in the
decay of squarks and the gluino which are hard enough to survive 
a strong $p_T^{jet}$ cut. 
One could exploit the jet-count (possibly supplemented by effective mass 
distributions)
to have a reasonably fair idea of the squark and the gluino masses. 
The low leptonic activity
compared to the jets then suggests VLSP sleptons as a
viable explanation.


\section{SSD's and New Physics}

SSD's are not only very characteristic for SUSY
but perhaps also for a scenario like the recently proposed Universal Extra
Dimensions \cite{Appelquist:2000nn} where they could be generated in a manner
analogous to the way they are in SUSY, even though the UED spectrum
does not contain any majorana fermion. 

In UED, the level 1 Kaluza-Klein (KK) excitations of the SM
particles (which are the so-called level `0' modes in KK language)
could all lie in the 
TeV range and mimic a SUSY spectrum except for their spins. For example,
there is a one to one correspondence between each and every scalar-fermion 
in SUSY to a level 1 KK-fermion (level 1 quarks and leptons), between a 
(fermionic) gaugino in SUSY to a level 1 excitation of an SM gauge boson,
\emph{viz.}, level 1 gluon($g_1$), $\gamma_1$, $Z_1$, $W^\pm_1$ etc.. 
Thus the UED spectrum looks similar in appearance to SUSY, which 
motivated some recent studies
\cite{Bhattacharyya:2005vm, Battaglia:2005zf,Bhattacherjee:2005qe} 
on its collider phenomenology vis-a-vis SUSY.

In this section, we discuss how cascades involving level-1 UED excitations
could turn out to be very similar to the corresponding ones in SUSY. 
This is
possible since in both SUSY and UED the couplings among different species
are related to their SM counterparts. With an inbuilt
discrete symmetry called $K$-parity the UED phenomenology 
resembles that of SUSY with conserved $R$-parity. Level 1
quarks ($q_1$) and the $g_1$ can be produced at the LHC much in the same way as
the squarks and the gluino in SUSY. The similarity in couplings, a similar spectrum and the 
imposed $K$-parity ensure analogous decay chains for these strongly interacting
particles. A $g_1$ produced in the hard scattering could undergo
a $q_1$ mediated 3-body decay to a pair of SM quarks and a $W_1^\pm$ much
in the same way as the gluino in SUSY decays into a chargino:
\[ g_1 \stackrel{\bar{q}_1^*} \longrightarrow \qprime \qbar W_1^\pm\  . \]
On the other hand, level 1 quarks could follow a similar path in their
cascades to that of the squarks in SUSY.
If a pair of same sign $q_1$ is produced in hard scattering, 
they may undergo the following 
electroweak cascade in a way similar to squarks in SUSY,
\[ q_1 \to \qbarprime W_1^\pm\ . \] 
In either case the $W_1^\pm$ could undergo the following decay
\[ W_1^\pm \stackrel{\ell_1^*}{\longrightarrow} \ell \ellbarprime
\gamma_1\ ,\]
again like a chargino does in SUSY.
Here $\gamma_1$ is the level 1 excitation of the SM photon (more precisely
the U(1) gauge boson of the SM, since the effective electroweak mixing angle at the 
first KK-level is known to be small). Thus, in close analogy to SUSY
cascades the level 1 gluon (again, a color octet vector boson) or a level 
1 quark (an electromagnetically charged fermion) in UED, 
when produced in pairs (or in association) at the LHC,
would lead to the SSD's. It does not take majorana fermions 
to end up with SSD's. 


Qualitatively much of SUSY phenomenology (at colliders)
could have a close correspondence to UED and these
two approaches could potentially fake each other. The difference in 
the spins for corresponding particles in SUSY and the UED is one
direct probe into what is being observed.
Recently this issue 
has been addressed \cite{Bhattacharyya:2005vm, Battaglia:2005zf, 
Bhattacherjee:2005qe} 
in the context of a 
future linear collider. There are
recent studies on probing the spins directly
\cite{Smillie:2005ar,adkkkm}. Below we will propose a possible
way to address this issue and perhaps largely solve the problem.

The framework used in Refs.\cite{Bhattacharyya:2005vm, Battaglia:2005zf} 
is the minimal version of the
UED where one makes a simplifying assumption that the boundary kinetic terms 
vanish at the cut-off scale \cite{Cheng:2002iz}.  
That assumption, however, restricts the spectrum such that the
splitting among different KK excitations may not be enough to result in
hard enough jets and/or leptons that could pass the stringent cuts at the LHC.
Thus, such a (minimal) version of the UED is not likely to show up at
the LHC. 

On the other hand, the boundary terms receive divergent contributions
thus requiring counter terms. 
As observed in Ref.\cite{Cheng:2002iz}, the finite parts of 
these counter terms remain undetermined and are free parameters of
the theory. These arbitrary extra degrees of freedom could be exploited to end
up with an unconstrained UED (UUED) scenario where the different KK excitations
are reminiscent of an unconstrained MSSM spectrum. In 
Ref.\cite{Cheng:2002iz} a close analogy has been pointed out between
these two scenarios and a similarity between the unknown soft masses of MSSM
and the unknown counter terms of the UED is drawn. A qualitatively similar
effect is discussed recently in the context of an explicit Lorentz violation
within UED by studying a subset of operators that leaves 4-D
Lorentz invariance intact while breaking it in the 
5-th dimension\cite{Rizzo:2005um}. Phenomenologically, these
provide us with a broad scenario where faking (between SUSY and UED) 
could be rather complete qualitatively. This is especially significant at the LHC where
the kinematic cuts would help rule out the minimal version of the UED
(MUED) scenario.  

Table 3 gives us an impression of how efficient such a faking could be by 
comparing the corresponding MSSM and the UUED
production cross sections and the cascade branching fractions 
for the particles involved.
The input MSSM parameters used are $\mone=325$ GeV, $\mtwo=375$ GeV,
$\mthree=494$ GeV, $\mu=1$ TeV, $\tan\beta=10$ leading to $\mgluino=600$ GeV,
$\mntrl1=322$ GeV, $\mntrl2=385$ GeV. The soft scalar masses are fixed at 1 TeV
while the $A$ parameters are somewhat arbitrarily taken to be 200 GeV 
to ensure the sfermion mixing does not become much of an issue. 
An unconstrained version of UED \cite{adkkkm-comphep} was then tuned in 
CalcHEP/CompHEP 
\cite{calchep-comphep} to have an identical spectrum. The effective cascade rates in different final states were
then calculated by folding the basic cross sections (without cuts)
with the respective branching 
fractions for both MSSM and the UUED keeping track of the combinatoric 
factors\footnote{A detailed comparison of the different final states at 
the generator 
level is rather involved and technically challenging (handling multibody final 
states with 8 to 10 particles in the final state with full matrix elements 
as is characteristic for the CalcHEP/CompHEP). Nonetheless, the basic production
cross sections and the branching fractions presented here are likely to provide
a fair estimate.}.

%
The parton-distribution function is set at $Q = m_{g_1}(\mgluino)$. 
%
%
\begin{table}[h!t!b!]
\begin{center}
{\small
\renewcommand{\arraystretch}{1.1}
\begin{tabular}{||c||c|c||}
\hline
\hline
 & & \\
  & {\bf MSSM} & {\bf U-UED} \\
 & & \\
\hline
\hline
{\bf Production} & & \\
{\bf Cross sections} & $\sigma_{\gluino \gluino}=4.51$ pb & $\sigma_{g_1 g_1}=65.95$ pb \\ 
 & & \\
\hline
\hline
 & & \\
%
& $\gluino \to q\qbarprime \chpm1=0.45 $ & $g_1 \to q\qbarprime W_1^\pm=0.45$\\
& $\gluino \to q \qbar \ntrl2=0.28 $ & $g_1 \to q \qbarprime Z_1=0.28$ \\  
& $\gluino \to q \qbar \ntrl1=0.27 $ & $g_1 \to q \qbarprime B_1=0.27$ \\  
{\bf Branching} & & \\
 \cline{2-3}  
{\bf Fractions} & & \\
 & $\chpm1 \to q \qbarprime \ntrl1 = 0.67 $ 
 & $W_1^\pm \to q \qbarprime B_1 = 0.18 $ \\
 & $\chpm1 \to \ell \nu \ntrl1 = 0.33 $  
 & $W_1^\pm \to \ell \nu B_1 = 0.82 $ \\
& & \\
 & $\ntrl2 \to q \qbar \ntrl1 = 0.94 $ 
 & $Z_1^\pm \to q \qbar B_1 = 0.22 $ \\
 & $\ntrl2 \to \ell \ellbar \ntrl1 = 0.04 $  
 & $Z_1^\pm \to \ell \ellbar B_1 = 0.39 $ \\
 & $\ntrl2 \to \nu \bar{\nu} \ntrl1 = 0.01 $ 
 & $Z_1^\pm \to \nu \bar{\nu} B_1 = 0.39 $ \\
 & & \\
\hline  
\hline  
{\bf Cascade} & & \\
{\bf Fractions} & & \\
\cline{1-1}
1-lepton & 0.248 & 0.385 \\
OS 2-lepton & 0.030 & 0.183 \\
SS 2-lepton & 0.011 & 0.068 \\
3-lepton & 0.003 & 0.081 \\
 & & \\
\hline  
\hline  
{\bf Cascade} & & \\
{\bf Rates} & & \\
\cline{1-1}
1-lepton & 1.12 pb & 25.39 pb\\
OS 2-lepton & 0.13 pb & 12.06 pb \\
SS 2-lepton & 0.05 pb & 4.48 pb \\
3-lepton & 0.014 pb & 5.34 pb \\
 & & \\
\hline
\hline
\end{tabular}
}
\end{center}
\vskip -10pt
\caption{\small Cross-sections for $\gluino$-pair and $g_1$ pair productions
in MSSM and UUED respectively for an identical spectrum (see text) along with 
relevant branching fractions in their cascades.}
\end{table}
%
%
We see from Table 3 that the production cross section for the $g_1$ pair in UUED
far exceeds (an order of magnitude) that 
for the $\gluino$-pair in MSSM for identical
spectra\footnote{We will get back to this issue later in this
  section.}, because the $g_1$ has spin 1 \cite{Smillie:2005ar,ued-hadcol}.
Next we find that the branching fractions of 
$\gluino$ and the $g_1$ to chargino and $W_1^\pm$ are identical. 
This reflects the known similarity of couplings found in
the MSSM and the UED, especially when the lighter chargino is predominantly
a $\widetilde{W}^\pm$ (which is the case here as 
$\mu (500 \; {\mathrm{GeV}}) >> \mtwo (300 \; {\mathrm{GeV}})$).
As we proceed with the cascade, a flipping of the leptonic and jet
branching fractions is noted
for the decaying chargino when compared to a decaying $W_1^\pm$. Actually,
the chargino branching fractions resemble the same for $W^\pm$ in the SM.
This is because its 3-body decay is mainly mediated by $W^\pm$, the sfermions
being almost decoupled for the purpose because of their heaviness.
In UED $W_1^\pm$, a pure $SU(2)$ adjoint state,
does not couple to $B_1$, the pure $U(1)$ state. Thus, the 3-body decay
of $W_1^\pm$ is solely mediated by heavy KK-fermions\footnote{This is best 
reflected in the decay width of $W_1^\pm$ which, in the present case, is
${\cal O}(10^{-10} \; \mathrm{GeV})$ and almost 6 orders of magnitude smaller
than the chargino width. The latter might get further enhanced as the
mediating $W^\pm$ gets closer to being on-shell.}. $B_1$ appears in the
secondary vertex and since it has only hyper-charge couplings to fermions 
the leptonic widths are larger than the jet ones. This is especially true
when the fermions involved are left-handed in nature since they couple
to $W_1^\pm$. 

Table 3 thus indicates larger event rates both in the leptonic and jet
final states for the UUED. 
While a suppressed hadronic branching fraction of $W_1^\pm$
somewhat compensates for an enhanced $g_1$-pair cross section it is always
a win-win situation for the leptonic final state. Hence, for a similar
spectrum a larger imbalance in lepton and jet events is expected in UUED.
However, unless we have information about the actual masses involved, 
it could be difficult to decide on 
one or the other candidate scenario solely by looking at
the event rates. The problem is thus very generic in nature and not limited
to SSD's. More specifically, almost any {\it given} channel could be made to be the
same for SUSY at UUED. But, most importantly, the patterns for several
signatures are different, so it should not be difficult to distinguish
them with sufficient data. Particularly the above-noted flipping of
leptonic and jet branching ratios, when normalized appropriately can
distinguish. 

In fact, it is likely that one can use the connection of spin to
production cross section to distinguish between SUSY and UUED if a
discovery that seems naively to be consistent with both should occur.
As pointed out earlier UUED enjoys a larger
cross section compared to MSSM for a given mass scale. Thus to
distinguish,
one has to associate the observed event rates to a mass scale
relevant to the process. Here we examine a simple way to do this and
find a very robust result. Presumably more careful definitions of the
mass scales will strengthen the result when there is data.
\begin{figure}[tbh]
\begin{center}
\hspace{-3.3cm} 
\includegraphics[width=13cm]{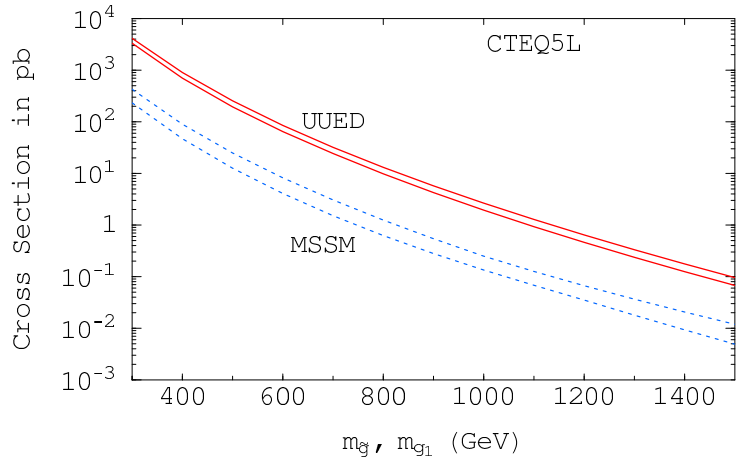}
\caption{Lowest order cross sections from CompHEP as a function of final 
state mass
for $p p \to \gluino \gluino (g_1 g_1)$ in the MSSM(UUED) 
at the LHC with $\sqrt{s}=14$ TeV. The bands are obtained by
varying the squark/level-1 quark masses and the 
renormalization/factorization scale $\mu$ over the range
$\frac{m_{\gluino(g_1)}}{2} \leq \mu \leq 2 m_{\gluino(g_1)}$. 
The CTEQ5L parton distribution is used. This shows that for a given
experimentally determined mass scale the cross sections are very
different, and are likely to allow distinguishing between such
interpretations of an LHC signal. }
\label{uued_mssm}
\end{center}
\end{figure}

As is well-known, a kinematic distribution related to the mass scale 
of new physics that could be studied at the LHC is the 
so-called effective mass \cite{Hinchliffe:1996iu}
\begin{eqnarray*}
M_{eff} = E_T^{miss} + \sum_{jet} p_{T, jet} .
\end{eqnarray*}
This peaks at around twice the mass scale defined naively as
\begin{eqnarray*}
M = min(m_{g^*},m_{q^*}) 
\end{eqnarray*}
where $g^*$ and $q^*$ are the excited gluon and quark states of 
SUSY or UED in our present study.
Thus for a given position of the peak 
of $M_{eff}$ a much larger event rate would occur for UUED than for
SUSY. A definitive conclusion would require that the resulting event
rates did not significantly overlap.

Fig. \ref{uued_mssm} illustrates the variation of cross sections for 
gluino-pair (level-1 gluon pair) production against
the mass scale which is the gluino (level-1 gluon) mass for the MSSM (UUED) 
with squarks (level-1 quarks) being much heavier.
In both cases the bands are
obtained by varying the squark/level-1 quark 
masses and the renormalization/factorization scale 
$\mu$ which have the strongest impact on the cross sections at the lowest
order. We note that these theoretical 
bands never
overlap over a significant range of the mass scale. Thus, a conclusion
favoring one or the other scenario should be rather
robust. Of course, one has to critically estimate
higher order effects on the theoretical side and check many of the experimantally
involved issues. There may be a better procedure than identifying 
the peak of the effective mass curve with the
mass scale. We expect that once
there is data fitted by SUSY and UED models, then sharper procedures
will easily be implemented. The basic point is that for given masses
the connection between spin and cross section is fundamental, and is a
tool that can remove any confusion. It becomes useful once data sets
the mass scales. Our comparison of UUED and SUSY was done at mass
scales that might already be excluded for UUED by the recent analysis
of Flacke, Hooper and March-Russell \cite{Flacke:2005hb}. 
Their improved analysis of precision EW data sets new lower
limits on the UED compactification scale. In the context of UUED these
limits might perhaps change, but even if they survive, our conclusions will
remain if we increase the mass scales.

\section{Conclusions}

Minimal SUGRA along with other prominant SUSY breaking 
scenarios have found enormous use in developing benchmarks 
for systematic studies both at theoretical and 
experimental levels. These are helping us to learn 
beforehand how to better prepare to recognize what is discovered, and
how to analyze it.
It is important to maintain a critical attitude 
towards any model dependent approach, so that it does not lead to 
oversimplification and overlooking other possibilities. 
One such possibility is the weakening of the leptonic signals at the LHC. 

Same-sign dilepton signals (SSD) are a
rather robust signature of SUSY. But one might wonder if the SSD signature
is a ``sufficient'' and ``necessary'' signature of SUSY.
There are situations when SSD's can 
get depleted to an unobservable level while (broken) SUSY could still be a
valid description of Nature (i.e. they are not ``necessary'').
Second, the existence of the SSD signal alone does not automatically exclude
alternative descriptions of nature, \emph{viz.} in the
form of Universal Extra Dimensions (UED) (they are not ``sufficient''
either). We analyze both of these situations.

Once we give up mSUGRA and start working in the MSSM,
depletion of SSD events may occur in several possible ways as 
discussed in the text.

Leptons in different leptonic channels tend to have the same
origin in the decay cascades. Thus typically all  
leptonic final states have decreased rates if SSD do,
though at different rates, dictated mainly by the combinatorics.
Also, we outline how some 
leptonic depleted modes can actually serve as a
powerful input for our understanding. 

Depletion in some channel(s) may not
mean reinforcement in some others. This is counter to naive expectations
on the grounds of conservation of normalizations of different
branching fractions down the cascades. An apparent violation
of such normalizations could show up when 
kinematic cuts affect preferably some particular final states.
Such apparent violations can be drastic when one of the 
channels has a low event rate and is on 
the verge of being unobservable while others have a 
rather healthy rate to start with.
If a Same-Sign dilepton signal is found, the default hypothesis is
that a signal for supersymmetry has been seen. Even in the full MSSM
parameter space there will be consistency checks with other
channels. If a signal is seen but it does not include Same-Sign
Dileptons, a SUSY interpretation
implies constraints on rates in other leptonic channels.

On the other hand, it might turn out that a completely different physics scenario 
(an unconstrained UED or UUED in our study) is behind a
particular signature (SSD, in our case) otherwise thought to be 
a ``smoking gun'' signature
for SUSY. 
The phenomenology of the UED at the hadron colliders
provides a rather concrete example of a scenario that could apparently
fake SUSY. 
Recently such situations have been studied for linear colliders 
\cite{Bhattacharyya:2005vm,Battaglia:2005zf, Bhattacherjee:2005qe}.
For the SUSY/UUED case we have noted that the close and basic connection
between spins and cross section at a given mass is likely to allow
distinguishing them at LHC, once data on the mass scales is available, 
and further study of patterns of relative rates in
different channels will also be helpful.

We conclude that once one has a signal to analyze, it is reasonable to
be optimistic that one will quickly be able to recognize whether it is
likely to be the discovery of supersymmetry, even at a hadron
collider. If it is, then the challenges of measuring the effective
Lagrangian at the collider scale, and of trying to probe the
underlying theory \cite{Binetruy:2003cy,Bourjaily:2005ja,Kane:2005az}
will begin.

\vskip 20pt
\noindent
{\large \bf Acknowledgments}
\vskip 10pt
\noindent
This work is supported by the US Department of Energy and the Michigan Center
for Theoretical Physics.
The authors thank Steve Mrenna for frequent discussions and help on specific 
usages of Pythia. GK thanks Liantao Wang and Nima Arkani-Hamed for
conversations on signatures. AD thanks Jean-Loic Kneur, K. Kong, 
Margarete M$\ddot{\mathrm u}$hlleitner, Monoranjan Guchait, 
Jean-Paul Guillaud, Torbj$\ddot{\mathrm o}$rn Sj$\ddot{\mathrm
  o}$strand and Peter Skands for numerous discussions 
on usages of different routines and softwares.

\end{document}